\documentclass[twocolumn]{aastex62}
\usepackage{graphicx}
\usepackage{color}
\usepackage{amsmath,amssymb}
\usepackage{times}
\begin{document}
	
	\correspondingauthor{Emily C. Cunningham}
    \email{eccunnin@ucsc.edu}
    \author{Emily C. Cunningham}
	\affiliation{Department of Astronomy \& Astrophysics,
    			University of California, Santa Cruz,
				1156 High Street,
				Santa Cruz, CA 95064, USA}
    \author{Alis J. Deason}
    \affiliation{Institute for Computational Cosmology, Department of Physics, University of Durham, South Road, Durham DH1 3LE, UK}
    \author{Constance M. Rockosi}
    \affiliation{Department of Astronomy \& Astrophysics,
    			University of California, Santa Cruz,
				1156 High Street,
				Santa Cruz, CA 95064, USA}
    \author{Puragra Guhathakurta}
    \affiliation{Department of Astronomy \& Astrophysics,
    			University of California, Santa Cruz,
				1156 High Street,
				Santa Cruz, CA 95064, USA}
    \author{Zachary G. Jennings}
    \affiliation{Department of Astronomy \& Astrophysics,
    			University of California, Santa Cruz,
				1156 High Street,
				Santa Cruz, CA 95064, USA}
    \author{Evan N. Kirby}
    \affiliation{California Institute of Technology, 1200 E. California Blvd., MC 249-17, Pasadena, CA 91125, USA}
    \author{Elisa Toloba}
    \affiliation{Department of Physics, University of the Pacific, 3601 Pacific Avenue, Stockton, CA 95211, USA}
    \author{Guillermo Barro}
    \affiliation{Department of Physics, University of the Pacific, 3601 Pacific Avenue, Stockton, CA 95211, USA}
    
 	\title{HALO7D I: The Line of Sight Velocities of Distant Main Sequence Stars in the Milky Way Halo}

	\begin{abstract}
		The Halo Assembly in Lambda-CDM: Observations in 7 Dimensions (HALO7D) dataset consists of Keck II/DEIMOS spectroscopy and Hubble Space Telescope-measured proper motions of Milky Way halo main sequence turnoff stars in the CANDELS fields. In this paper, we present the spectroscopic component of this dataset, and discuss target selection, observing strategy, and survey properties. We present a new method of measuring line-of-sight (LOS) velocities by combining multiple spectroscopic observations of a given star, utilizing Bayesian hierarchical modeling. We present the LOS velocity distributions of the four HALO7D fields, and estimate their means and dispersions. All of the LOS distributions are dominated by the ``hot halo": none of our fields are dominated by substructure that is kinematically cold in the LOS velocity component. Our estimates of the LOS velocity dispersions are consistent across the different fields, and these estimates are consistent with studies using other types of tracers. To complement our observations, we perform mock HALO7D surveys using the synthetic survey software Galaxia to ``observe'' the Bullock \& Johnston (2005) accreted stellar halos. Based on these simulated datasets, the consistent LOS velocity distributions across the four HALO7D fields indicates that the HALO7D sample is dominated by stars from the same massive (or few relatively massive) accretion event(s).
	
	\end{abstract}
    
	
	\section{Introduction}
 
When a dwarf galaxy falls in to the Milky Way (MW) potential and is tidally disrupted, its stars become members of the MW stellar halo. The orbital timescales of these stars are long compared to the age of the Galaxy; thus, long after these debris have lost their spatial association, they remain linked by their kinematic (and chemical) properties. Six dimensional (6D) phase-space information and chemical abundances of halo stars can therefore be used to unravel the accretion events that have contributed to the mass assembly of the MW. With the Halo Assembly in Lambda-CDM: Observations in 7 Dimensions (HALO7D) survey, we are measuring 3D kinematic information and chemical abundances (as well as constraints on 3D positions) for distant main sequence (MS) MW halo stars.

The HALO7D dataset consists of Keck II/DEIMOS spectroscopy and Hubble Space Telescope (\textit{HST}) measured PMs measured of distant ($D \sim 10$--100~kpc) MW main sequence turnoff (MSTO) stars. In this paper, the first in the HALO7D series, we present the spectroscopic component of this dataset. In a companion paper (Cunningham et al. 2018b, in preparation; hereafter, Paper II), we present the PM dataset and analysis of the 3D kinematic sample.

This paper is organized as follows. In section \ref{sec:ms_intro}, we motivate the HALO7D survey, and place our survey in context with other MW halo studies. In section \ref{sec:vr_mot}, we introduce \textsc{Velociraptor}, our hierarchical Bayesian method for measuring the LOS velocities for our faint targets. In Section \ref{sec:Dataset}, we describe the HALO7D fields, target selection and observations. In Section \ref{sec:vel}, we present the details of the \textsc{Velociraptor} method. In Section \ref{sec:results}, we present the LOS velocity distributions for the four HALO7D fields, estimate their velocity dispersions, and compare our results with those derived from other tracers. In Section \ref{sec:bj}, we compare our resulting LOS velocity distributions with predictions from simulations. We summarize our findings in Section \ref{sec:concl}. 

\subsection{HALO7D: A Deep, Pencil Beam Complement to Gaia}
\label{sec:ms_intro}
	Our current picture of the stellar halo has largely been shaped by its giant population. Giants and evolved stars, such as red giant branch (RGB) stars, blue horizontal branch (BHB) stars, and RR Lyrae variables, have many advantages as halo tracers, particularly because of their bright absolute magnitudes. Giants have enabled the mapping of the stellar halo out to great distances: \cite{Slater2016} used K-giants to measure the density profile out to 80 kpc, \cite{Hernitschek2018} measured the density profile of the MW stellar halo out to 150 kpc with RR Lyrae from Pan-STARRS1, and \cite{Deason2018a} used BHBs in the Hyper Suprime-Cam survey to measure the density profile out to $\sim$200 kpc. These tracers have also revealed a wealth of substructure in the distant halo (see \citealt{Sesar2017} and \citealt{Conroy2018} as some recent examples). 
    
    Until recently, our kinematic knowledge of the stellar halo beyond $D\sim 10$ kpc has been limited to one component of motion (the line-of-sight (LOS) velocity) for these bright tracers. While progress has been made on measuring the LOS velocity dispersion profile (e.g., \citealt{Xue2008} with SDSS BHBs;
    \citealt{Cohen2017} with RR Lyrae), there has been little knowledge of the tangential motion of these stars until this year --- from the second \textit{Gaia} mission data release (\citealt{Gaia2018}). Increasing our knowledge of the tangential motions of stars, in order to better map our Galaxy and understand its formation and structure, is the primary science goal of the \textit{Gaia} mission (\citealt{Perryman2001}).
    
Giants and evolved stars represent the upper echelon of a stellar population. While they are bright, they are very rare: MS stars are the dominant population in every stellar population. In addition, it is difficult (perhaps impossible) to uniformly select giants across all age and metallicity populations in the halo. For example, RR Lyrae only can evolve in metal poor populations, while M giants are only found in metal rich populations (see \citealt{Price-Whelan2015} for a discussion on how the relative numbers of RR Lyrae and M Giants in a population can be used to constrain the metallicity of a progenitor). While MS stars are fainter than giants, they are also more numerous, and all populations, regardless of age or metallicity, contain MS stars.
    
    While MS stars are ideal tracers thanks to their presence in all stellar populations, they are challenging to observe, because they are faint. Because of its limiting magnitude of $G\sim 20$,
    beyond $D\sim 15$ kpc in the halo, \textit{Gaia} will not provide proper motions for distant halo MS stars. The only instrument presently capable of measuring the proper motions (PMs) of distant ($D>20$ kpc) MW MS stars is \textit{HST}. \textit{HST} is a powerful instrument for precision astrometry, due to its stability, high spatial resolution and well-studied geometric distortions and point spread functions (PSFs) (e.g., \citealt{Anderson2006}). Multi-epoch \textit{HST} imaging has been exploited to make extremely accurate PM measurements of resolved stellar systems in the Local Group (LG), including the Magellanic Clouds (\citealt{Kallivayalil2006b}, \citeyear{Kallivayalil2006a}, \citeyear{Kallivayalil2013}), MW globular clusters (\citealt{Sohn2018}), MW dwarfs Draco and Sculptor (\citealt{Sohn2017}), and M31 (\citealt{Sohn2012a}).
    
    The first individual MW stars with measured \textit{HST} PMs were published by \cite{Deason2013b} (hereafter D13). These faint stars ($21<m_{F606W}<24.5$)
    had their PMs measured serendipitously during the \cite{Sohn2012a} M31 PM study. The third component of the motion for these stars, the LOS velocity, was measured by \cite{Cunningham2016}, using the DEIMOS spectrograph on the Keck II telescope, making this sample of 13 stars the first sample of stars with measured 3D kinematics outside the solar neighborhood.
D13 and C16 confirmed that we can measure kinematic properties of distant MS stars with these two world class telescopes. However, these studies were limited to only 13 stars across three \textit{HST} pointings. More stars and lines of sight through the halo are required to use MS star kinematics to investigate the formation of the Galaxy.
	
	The HALO7D survey aims to address the current lack of distant MS stars with measured 3D kinematics. This dataset is unique even in the era of \textit{Gaia}, measuring 6D phase space information for MS stars as faint as $m_{F606W}\sim 24.5$. In order to obtain spectra of these stars with sufficient signal-to-noise (S/N) for LOS velocity and abundance measurements, deep spectroscopy with a large telescope is required. HALO7D complements the \textit{HST} proper motions with deep spectra (8--24 hour integrations) observed with Keck II/DEIMOS. However, spectra of individual stars at this depth is unprecedented, and new techniques are required to make measurements with these data. For the interested reader, we motivate this new technique in the next subsection; for those interested in the survey details, please skip ahead to Section \ref{sec:Dataset}.

	\subsection{The Need for \textsc{Velociraptor}: Challenges of Deep Slit Spectroscopy}
    \label{sec:vr_mot}
 
  \textit{HST} can measure proper motions for exceedingly faint stars. The deep spectroscopy required to observe to these same magnitudes presents a significant challenge! To achieve the depth of our survey, targets were observed over multiple nights, and, in some cases, over years. In order to combine our different spectroscopic observations of a given target into a single measurement of the star's velocity, we required a new approach that took into account the fact that different observations of the same star will have different velocities. The \textsc{Velociraptor} software employs Bayesian hierarchical modeling in order to combine multiple, often noisy, observations of a star, each with different zero-point offsets, into a single posterior probability distribution for the star's velocity.
    
    One origin of zero-point offset is slit miscentering. Because stars are point sources, they do not fill the full width of the slit during observations (thanks to the exquisite seeing on Mauna Kea). If the star is not perfectly centered 
    in the slit, the wavelength solution for the object is slightly offset from the wavelength solution given by the calibration of arc lamps
    (e.g., \citealt{Sohn2007, Simon2007}). This wavelength solution difference corresponds to an apparent velocity shift that can be measured from the velocity of the telluric A-band absorption feature. Velocities of telluric features should be $0$ km~s$^{-1}$
    if the wavelength solution is correct. We refer to this velocity offset as the \textit{A-band correction} ($v_{\rm Aband}$)\footnote{While it has not been treated as such in the literature, we note that the effect of this slit miscentering is closer to a wavelength shift than a velocity shift. In this work, we measure stellar velocities using both the H$\alpha$ and Ca triplet absorption features. Because these absorption features are approximately equidistant in wavelength from the telluric A-band feature, we can safely treat the A-band correction as a velocity offset, as others have done in the past. However, if one were to measure velocities using \textit{only} H$\alpha$ or Ca, or if one were to also use lines farther in the blue or red, then it is better to treat the A-band correction as a wavelength shift.}, and it is subtracted from the raw velocity ($v_{\rm raw}$, the velocity of stellar absorption features in an \textit{observed} spectrum),
along with the heliocentric correction ($v_{\rm helio}$, due to the Earth's motion around the Sun) to yield the corrected velocity in the heliocentric frame:

		\begin{equation}
			v=v_{\rm raw}-v_{\rm Aband}-v_{\rm helio}.
		\end{equation}

		The offset in the slit can be due to astrometric errors or slight mask misalignment. As such, the A-band correction varies from object to object on a given mask, and varies from observation to observation of the same object. For a star observed through a 1 arcsecond slit with Keck II/DEIMOS, configured with the 600ZD grating, $v_{\rm Aband}$ can be up to $\approx \pm 60$ km~s$^{-1}$.
Given that the velocity dispersion of the halo is on the order of $100$ km~s$^{-1}$, and that velocity dispersions of streams and dwarfs can be less than 10 km~s$^{-1}$, it is essential to take into account this velocity offset before combining spectra from different observations. In addition, when the spectra in question are noisy, the measurements of $v_{\rm Aband}$ are also noisy, and their uncertainties need to be incorporated in to the ultimate measure of the corrected velocity uncertainty. In order to best leverage our signal to get accurate velocity measurements of our stars, along with correct uncertainties, we employ Bayesian hierarchical modeling to measure the velocities of individual spectroscopic measurements and their differing zero-point offsets \textit{simultaneously}.   
        
        For the details of the model implemented by \textsc{Velociraptor}, we refer the reader to Section \ref{sec:vel}. Fake data testing and further details are discussed in the Appendix.

	\section{Data: The HALO7D Keck Program}
	\label{sec:Dataset}
    
    In this section, we describe the properties of HALO7D Keck II/DEIMOS spectroscopic program. We describe our choice of survey fields in Section \ref{sec:fields}; our target selection procedure is outlined in Section \ref{sec:targ}; and observations are described in Section \ref{sec:obs}. The extragalactic ``piggy-back'' programs are briefly described in \ref{sec:eg}. In Section \ref{sec:samp}, we describe how we selected the halo star candidates used for dynamical modeling from the spectroscopically observed targets.
	
	\subsection{Survey Fields}
    \label{sec:fields}
	
	For a 3D kinematic study of distant halo stars, we aimed to survey high latitude fields that were characterized by many, contiguous, multi-epoch \textit{HST} pointings. Deep, multi-epoch imaging is required in order to measure proper motions of distant main sequence stars (D13), while the large field of view of DEIMOS enables efficient spectroscopic follow-up of many contiguous \textit{HST} pointings. The fields targeted by the Cosmic Assembly Near-infrared Deep Extragalactic Legacy Survey (CANDELS; \citealt{Grogin2011}; \citealt{Koekemoer2011}; PIs: S. Faber, H. Ferguson) were therefore a natural choice. The \textit{HST} footprints of the CANDELS fields were designed with spectroscopic follow-up in mind,  and they have been observed by \textit{HST} many times over the course of \textit{HST}'s operation. 
	
\begin{figure*}
	\includegraphics[width=\textwidth]{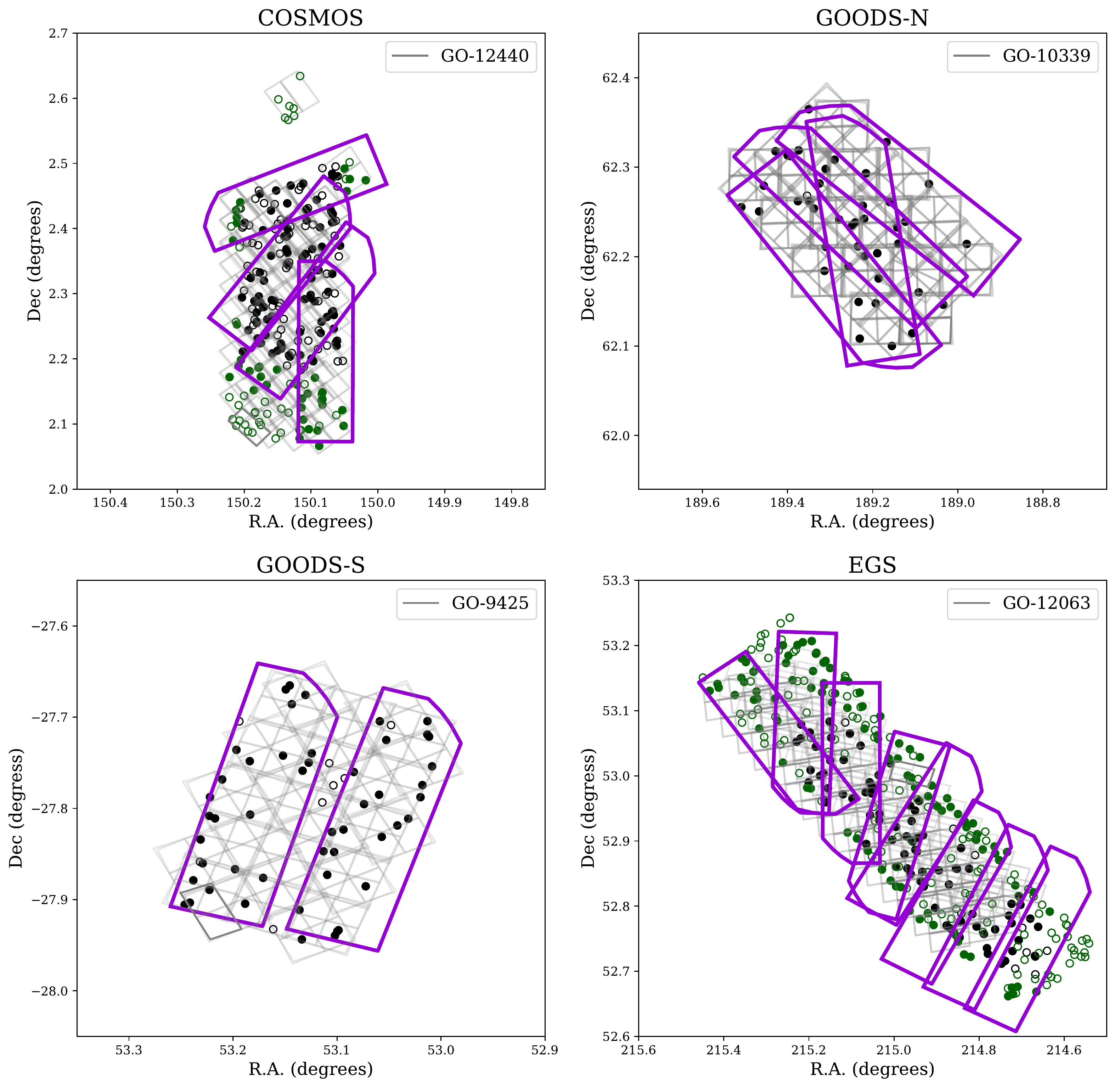}
	\caption{Footprints of the four HALO7D fields. Black points are the positions of halo star candidates selected from the CANDELS catalogs. Filled points were observed with DEIMOS, while empty circles denote halo star candidates that were not observed. Green points in EGS and COSMOS are halo star candidates selected from the IRAC and UltraVista catalogs, respectively. DEIMOS mask outlines are drawn in purple. Grey squares indicate one epoch of HST imaging that is used for measuring PMs of these same stars (see Paper II).}
	\label{fig:footprints}
\end{figure*}

	HALO7D surveyed four out of the five CANDELS fields: EGS, COSMOS, GOODS-N, and GOODS-S (the fifth field, UDS, only has one epoch of \textit{HST} imaging). The coordinates of the four HALO7D fields are listed in Table \ref{tab:fields}; their relatively high latitudes, resulting in minimal foreground contamination from MW disk stars, makes them ideal for both extragalactic and MW halo studies. 
    Figure \ref{fig:footprints} shows the footprints of the four HALO7D fields. Tiling patterns for one epoch of HST imaging are shown in grey; the HALO7D Keck/DEIMOS mask pointings are shown in purple (see Section 2.3). 

\begin{table*}
	\begin{center}
	\begin{tabular}{c  c  c  c  c c c }
		Field & R.A. (J2000) & Dec (J2000) & \textit{l} (deg) & \textit{b} (deg) & Area (arcmin$^2$) & Catalog Refs.\\
		\hline \hline
		COSMOS & 10:00:28 & +02:12:21 & 236.8 & 42.1 & 288  & \cite{Nayyeri2017}, \cite{Muzzin2013}$^{\dagger}$\\
		GOODS-N & 12:36:44 & +62:14:24 &125.9 & 54.8 & 166 & Barro et al. (2018; in prep)\\
		GOODS-S & 03:32:30 & --27:48:11 & 223.6 & -54.4 & 160 & \cite{Guo2013} \\
		EGS & 14:15:29 & +52:08:19 & 96.4 & 60.4 & 384 & \cite{Stefanon2017}, \cite{Barro2011}$^{\dagger}$\\
	\end{tabular}
	\end{center}
	\caption{Coordinates of the four CANDELS fields studied in HALO7D. These fields were chosen for their deep, multi-epoch \textit{HST} photometry. The listed field area corresponds to the field area covered with multi-epoch imaging. Catalogs indicated with daggers were used in the secondary target selection; see Section \ref{sec:add_targ}.}
	\label{tab:fields}

\end{table*}

	\subsection{Halo Star Candidate Selection}
    \label{sec:targ}
	
	Halo star candidates were selected using $u$, F606W (broad \textit{V} filter) and F814W (broad \textit{I} filter) photometry, from the catalogs listed in Table \ref{tab:fields}. Star candidates were identified based on image morphology (using the SExtractor parameter \verb|class_star|; \citealt{Bertin1996}) measured in WFC3 F160W images (the images used for source detection for CANDELS; see catalog references in Table \ref{tab:fields} and references therein for more details on source detection and photometry). In order to select as many stars as possible, we used the fairly generous stellarity cut of \verb|class_star|$>0.5$ (a more typical stellarity threshold for a study interested in including as many galaxies as possible would require that stars have \verb|class_star|$>0.98$). We also excluded all targets which have non-zero measured redshifts. 
	
\begin{figure}
	\includegraphics[width=0.5\textwidth]{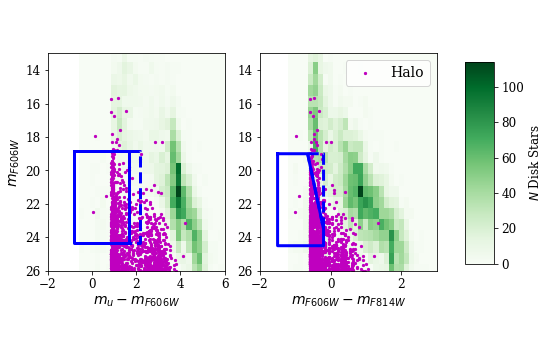}
	\caption{CMDs for stars in the Besan\c{c}on Galaxy Model, in a 1 deg$^2$ field of view centered on the coordinates of EGS. The green density maps show the CMD locations of the disk stars, with the number of disk stars in each CMD bin indicated by the colorbar. Halo stars are shown in magenta; only one out of five halo stars are shown for clarity. The HALO7D selection boxes are shown in blue. Stars were assigned priority based on their positions in these two CMDs: stars were assigned top priority if they fell within both solid selection boxes, and lowest priority if they fell into only one of the dashed selection boxes.}
	\label{fig:bes_cmds}
	\end{figure}

	To determine our selection boxes for optimally selecting halo stars, we used the Besan\c{c}on Galaxy Model (\citealt{Robin2003}). Figure \ref{fig:bes_cmds} shows color magnitude diagrams (CMDs) generated from the Besan\c{c}on model from a 1-square degree field centered on the coordinates of the EGS field. Green points are disk members and halo stars are shown in magenta. In order to target as many halo star candidates as possible with minimal disk contamination, we targeted faint, blue stars. Figure \ref{fig:bes_cmds} shows the HALO7D selection boxes in blue; our highest priority selection boxes are shown with solid lines, and the dashed line indicates our lower priority selection box.  
	
	Figure \ref{fig:h7d_cmds} shows the CMDs for the four HALO7D fields, and our selection boxes in blue. Magnitudes in the F606W and F814W bands are in the STMAG system. The $u$ band photometry is from ground based imaging, and magnitudes are in the AB system. In COSMOS and EGS, we used CFHT $u$ band photometry; in GOODS-S, we used CTIO $U$ band photometry; and in GOODS-N, we used KPNO $U$ band photometry (see references in Table 1). 
    
    Targets were assigned a priority for selection on a scale from 1-4 (with 4 being highest priority) based on the selection boxes:
	
	\begin{itemize}
		\item{Priority 4: Target falls in both solid selection boxes.}
		\item{Priority 3: Target falls in one of the solid boxes.}
		\item{Priority 2: Target falls in both dashed boxes.}
		\item{Priority 1: Target falls in one of the dashed boxes.}
	\end{itemize}
	
	\subsubsection{Additional Target Selection}
    \label{sec:add_targ}
	
	In the COSMOS and EGS fields, the CANDELS catalogs (developed from WFC3) did not overlap the full area with multi-epoch ACS imaging. To select targets in these regions (which contain stars that can have measured PMs), we used additional catalogs. In EGS, we used the ACS F606W/F814W fluxes published in the \cite{Barro2011} photometric catalog, and used the same prioritization scheme as described above. Sources in this catalog were identified using IRAC 3.6+4.5 $\mu$m imaging. Stars were identified by combining eight stellarity criteria based on photometric and morphological properties; see section 3.1 of \cite{Barro2011} for more detail. We included all targets with a total sum of stellarity criteria greater than 2, meaning that it was classified as star-like by at least two of the eight stellarity criteria (greater than 3 would be typical, but we again made our selection generous in the interest of not excluding stars). 
	
\begin{figure*}
	\includegraphics[width=\textwidth]{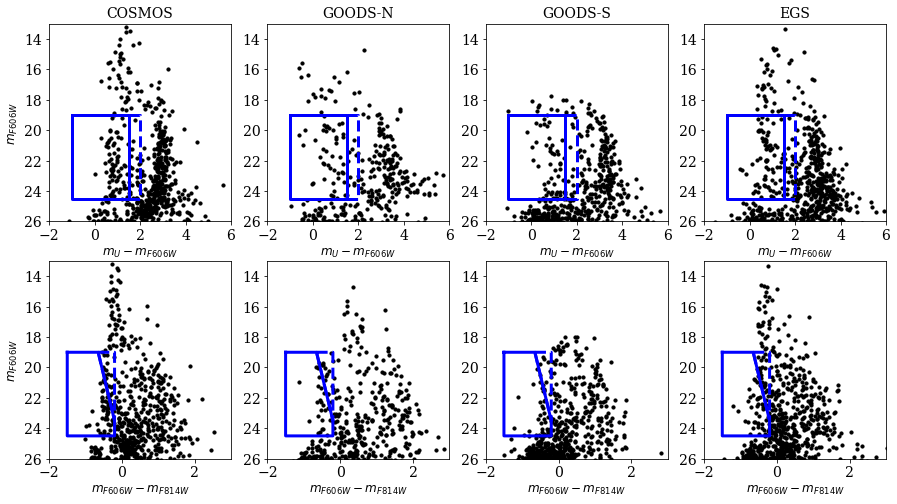}
	\caption{CMDs for stars in the four HALO7D fields. Selection boxes are shown in blue. Stars were assigned priority based on their positions in these two CMDs: stars were assigned top priority if they fell within both solid selection boxes, and lowest priority if they fell into only one of the dashed selection boxes. Magnitudes in F606W and F814W are computed in the STMAG system. We note that the bright stars in GOODS-S have been masked out in the catalog used for selection.}
	\label{fig:h7d_cmds}
	\end{figure*}

	In COSMOS, we selected targets from the $K_s$ selected catalog of the COSMOS/UltraVISTA field from \cite{Muzzin2013}. Stars are classified in this catalog using $u^*-J, J-K_s$ colors; see section 3.3 and Figure 3 in \cite{Muzzin2013}. To select HALO7D targets, we used the CFHT $u$ band and the Subaru $V$ band fluxes (\citealt{Capak2007}), using the same selection box as in the top panels of Figure \ref{fig:h7d_cmds}.
	
    \subsection{Extragalactic Targets}
    \label{sec:eg}
    
    Because the stellar halo is so diffuse, we typically placed only $\sim 25$ halo star candidates on a given DEIMOS mask. DEIMOS masks can contain up to $\sim 150$ slits: this provided an opportunity to obtain deep spectra for extragalactic targets as well as Galactic targets. These data have been used to study galactic winds in $z\sim1$ (\citealt{Yesuf2017}, Yesuf et al., in prep); quiescent galaxies at $z\sim 0.7$ (Conroy et al., in prep); internal galaxy kinematics (Wang et al., in prep); and dwarf galaxies (Guo et al., in prep).
    
	\subsection{Observations}
    \label{sec:obs}
	Spectra were obtained on the Keck II telescope with the DEIMOS spectrograph (\citealt{Faber2003}). Observations took place over the course of three years, beginning in March 2014 and ending in April 2017. While this program was intended to be completed over 19 nights in three semesters of observing, due to poor weather, observations extended through four spring semesters of observations plus several fall nights. 
    
	Observations were conducted with the same DEIMOS configuration as described in C16, and we summarize the key details here. For HALO7D observations, DEIMOS was configured with the 600 line/mm grating centered 7200 \AA, resulting in a typical wavelength range of 5000-9500 \AA. In the interest of limiting flux losses due to atmospheric dispersion, we divided our nights into observing ``blocks" of 1-2 hours each, and tilted the slits on our masks so that their position angles were consistent with the median parallactic angle of the observing block. Our typical exposure time was 20 minutes.
	
\begin{figure*}
	\includegraphics[width=\textwidth]{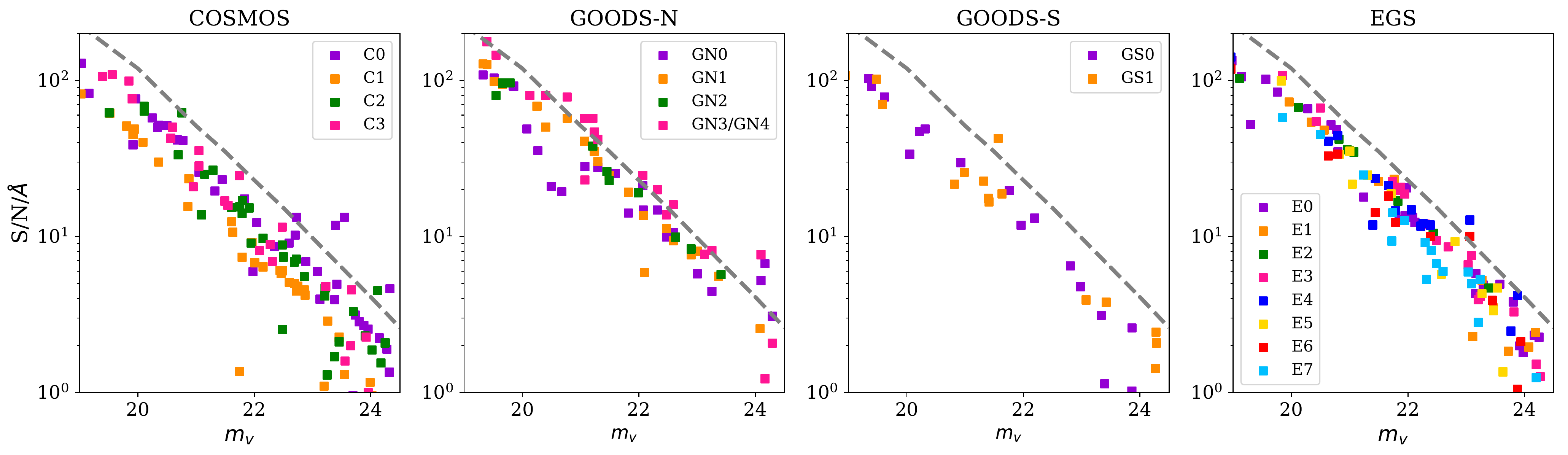}
	\caption{Signal to noise per angstrom for all HALO7D masks, as a function of $V$-band magnitude. Grey dashed lines indicate the predicted signal to noise with 8 hours of exposure time.}
	\label{fig:snr}
	\end{figure*}

	Our goal was to expose each mask for 8 ``effective'' hours: we sought to achieve the signal to noise as a function of apparent magnitude predicted by the DEIMOS exposure time calculator for 8 hours of exposure (grey dashed lines in Figure \ref{fig:snr}). Signal to noise (computed at $H\alpha$) for each mask as a function of apparent magnitude is shown in Figure \ref{fig:snr}; in practice, we achieved a typical $\sim 5-6$ effective hours of exposure on most masks.  
	
	DEIMOS mask footprints are shown on top of HST pointings in each of the four fields in Figure \ref{fig:footprints}. We observed eight masks in EGS, four masks in GOODS-N, two masks in GOODS-S, and four masks in COSMOS; properties of our observed masks are listed in Table \ref{tab:masks}. For one of our mask pointings in GOODS-N, we observed the same mask for twice as long as the other masks, but switched the list of extragalactic targets after an effective 8 hours was reached (GN3/GN4 masks have same pointings and MW target lists, but different extragalactic targets).   
	
	The slitmasks were then processed by the \textit{spec2d} pipeline developed by the DEEP2 team at UC Berkeley (\citealt{MCooper2012}). Table \ref{tab:sample} summarizes the progression of the HALO7D sample, from all CMD-identified halo star candidates to stars used for kinematic analysis in the subsequent sections. As seen in Figure \ref{fig:footprints}, we weren't able to observe all of our CMD selected candidates; this is reflected in the difference between the columns 2 and 3 of Table \ref{tab:sample}. In addition, as with any observational program, we suffered the occasional loss due to errors in the reduction (e.g., the object is too near the edge of the mask, bad columns, etc.; column 6 of Table \ref{tab:sample}). There were also targets that did not achieve sufficient S/N in their spectra for a measured velocity (column 7 of Table \ref{tab:sample}). Finally, there were also contaminants to our sample of CMD-selected MS stars that we identified spectroscopically; we discuss these contaminants in the next subsection.

	\subsection{Spectroscopically Confirmed Contaminants}
    \label{sec:samp}
	
	Following observations, all successfully reduced spectra were visually inspected in order to identify obvious contaminants in the MSTO sample.
    There were two sources of contamination that were identified spectroscopically and removed from the sample: extragalactic contaminants (column 4 of Table \ref{tab:sample}) and Galactic disk contaminants (column 5 of Table \ref{tab:sample}). 
	
	Extragalactic contaminants include quasars and emission line galaxies; our galaxy contamination rate was the highest amongst targets selected from the \cite{Barro2011} catalog, for which we had the most generous stellarity cut. 
	
	Our spectroscopically confirmed Galactic disk contaminants are white dwarf (WD) stars and red disk stars. We identified two types of WDs in our sample; we found WDs with very broad Balmer features as well as WDs with strong continua but no absorption features (these objects have disk-like PMs; see Paper II). The red disk stars contain obvious titanium oxide features in their spectra. The red stars in our sample made it into our selection boxes because they are located on the sky close to blue galaxies, which resulted in blended colors for the ground-based $u$ band photometry used for target selection. While we model contamination from blue disk main sequence stars in Section \ref{sec:results}, we exclude the WD and red stars from our sample for dynamical modeling. 
	    In the following Section, we describe in detail how we measure LOS velocities for our target spectra. To skip straight to the results, we refer the reader to Section \ref{sec:results}.
	
\begin{table*}
	\begin{centering}
	\begin{tabular}{c  c  c c c c c c c}
		Field & Mask Name & R.A. (J2000) & Dec (J2000) & Mask P.A. (deg) & Semesters Observed & MW Targets & Extragalactic Targets \\
		\hline \hline
		COSMOS \\
		\hline
		& C0 & 10:00:36.50 & $+$02:20:47.8 & $-38.2$ & 2014A, 2015A & 39 & 83\\
		& C1 & 10:00:31.52 & $+$02:16:14.6 & $-36.2$ &2017A & 38 & 80\\
		& C2 & 10:00:23.41 & $+$02:11:54.4 & $-0.20$ &2017A & 38 &74 \\
		& C3 & 10:00:29.45 & $+$02:26:16.0 & $111.8$ &2016A &23 &81\\
		\hline
		GOODS-N \\
		\hline
		& GN0 & 12:37:08.33 & $+$62:12:44.6 & $-142.0$ & 2015A, 2016A& 23 & 95\\
		& GN1 & 12:37:01.22 & $+$62:14:05.3 & $46.1$ & 2014A, 2015A& 24 & 105\\
		& GN2 & 12:36:38.82 & $+$62:15:48.8 & $51.3$ & 2016A &12 &95\\
		& GN3,GN4 & 12:36:58.73 & $+$62:13:02.6 & $9.4$ & 2014A, 2015A, 2016A &23, 21&104, 94\\
		\hline
		GOODS-S \\
		\hline
		& GS0 & 03:32:18.81 & $-$27:49:04.9 & $-17.3$ & 2015B, 2016B & 25 & 92, 88, 90\\
		& GS1 & 03:32:47.28 & $-$27:47:26.8 & $-15.6$ & 2015B, 2016B & 24 & 85, 85, 85\\
		\hline
		EGS \\
		\hline
		& E0 & 14:20:15.98 & $+$53:01:13.9 & $180.0$ & 2014A& 30 &101\\
		& E1 & 14:18:48.21 & $+$52:45:18.4 & $-26.1$ & 2014A& 19&103\\
		& E2 & 14:19:33.70 & $+$52:49:40.4 & $-28.1$ & 2014A&24 &97\\
		& E3 & 14:19:51.37 & $+$52:55:04.1 & $-30.4$ & 2015A& 28&81 \\
		& E4 & 14:20:42.48 & $+$53:05:50.4 & $177.8$ & 2015A&34 &80\\
		& E5 & 14:19:10.83 & $+$52:47:16.3 & $-25.8$ & 2016A& 22&81 \\
		& E6 & 14:21:03.34 & $+$53:04:42.3 & $-144.7$ & 2017A& 29&87\\
		& E7 & 14:19:49.96 & $+$52:56:05.3 & $164.0$ & 2017A& 31&79\\
	\hline\hline
	\end{tabular}
	\end{centering}
	\caption{Summary of the masks observed through HALO7D. In GOODS-N, the masks GN3 and GN4 contain the same MW targets, but different extragalactic targets. In GOODS-S, three sets of extragalactic targets were observed alongside each mask of MW targets.}
	\label{tab:masks}

\end{table*}

\begin{table*}
	\begin{center}
	\begin{tabular}{c  c  c c c c c c c}
		Field & $N_{\mathrm{CMD}}$ & $N_{\mathrm{Obs}}$ & $N_{\mathrm{Gal}}$ & $N_{\mathrm{WD}}$ & $N_{\mathrm{Red Stars}}$ & $N_{\mathrm{Reduction Errors}}$  & $N_{\mathrm{Low S/N}}$ & N Halo Star Candidates\\
		\hline \hline
		COSMOS & 101,67 & 87,36& 5,2 & 1 & 2& 4 & 21 & 88\\
		GOODS-N & 48 & 47 & 1 & 2&4 & 1 & 6 & 33 \\
		GOODS-S & 57 & 49 & 8 & 2&1 & 2 & 11 & 25   \\
		EGS & 96,135 & 84,94 & 6,28 & 7&7 & 8 & 25 & 97\\
        \hline\hline
	\end{tabular}
	\end{center}
	\caption{Summary of the progression of the HALO7D sample, from CMD selected targets to objects used in kinematic analysis. In COSMOS and EGS, we first indicate halo star candidate selected from the CANDELS catalogs, followed by candidates from the secondary catalogs for the first three columns of the table (see Section \ref{sec:add_targ}).}
	\label{tab:sample}

\end{table*}
	
	\section{Hierarchical Bayesian LOS Velocities: Velociraptor}
	\label{sec:vel}
	
	\begin{table*}
		\begin{centering}
		\begin{tabular}{c c c c }
			Level &  Parameters & Prior & Description\\
			\hline \hline
			Spectral Region & $\theta_{\rm Line}$ & \\
			\hline
			 & $v_{\rm Line}$ & $p(v_{\rm Line})\propto$ const &Velocity of region\\
			 &$C$ & $C+1 \sim $Gamma(2, 2)& Absorption line strength parameter\\
			 &$b_l$ & $p(b_l)\propto$ const& Legendre polynomial coefficients for continuum\\
			\hline
			Single Observation with J Regions& $\theta_{\rm spec}$ & \\
			\hline
					 & $v_{\rm raw}$ & $p(v_{\rm raw}) \sim \mathrm{Unif}[-600, 600]$ &Velocity of stellar absorption regions (e.g.,$H\alpha$)\\
					 & $v_{\rm Aband}$ & $p(v_{\rm Aband})\sim \mathrm{Unif}[-100,100]$ &Velocity of telluric region(s)\\
					 &$C_1, ..., C_J$ & $C_j+1 \sim $Gamma(2, 2)& Absorption line strength parameters\\
					 &$b_{l,1}, ..., b_{l,J}$ & $p(b_{l,j})\propto$ const& Legendre polynomial coefficients for continuum\\
			\hline
			Star with K Observations & $\theta$ & \\
			\hline
			 		&$v$ & $p(v)\sim \mathrm{Unif}[-600,600]$ & LOS velocity of the star \\
					&$\sigma_v$ & $p(\sigma_v) \sim \mathrm{Inv-Gamma}(7,72)$ & Dispersion of measurements; systematic error \\
					&$v_{\mathrm{raw},k},v_{\mathrm{Aband},k} $ & $v_{\rm{corr}}=v_{\mathrm{raw},k}-v_{\mathrm{Aband},k}-v_{\mathrm{helio},k}$ & Raw and A-band velocities of individual spectra\\
					& & $v_{\mathrm{corr},k} \sim \mathrm{N}(v, \sigma_v^2)$ & \\
					 &$C_{1,1}, ..., C_{J,K}$ & $C_{j,k}+1 \sim $Gamma(2, 2)& Absorption line strength parameters\\
					 &$b_{l,1,1}, ..., b_{l,J,K}$ & $p(b_{l,j,k})\propto$ const& Legendre polynomial coefficients for continuum\\
			\hline
		\end{tabular}
		\end{centering}
		\caption{Summary of model parameters (and priors) for different levels of our hierarchical model. In the first level of the model, we model a region of the spectrum containing an absorption line feature, such as H$\alpha$, the telluric A-band region, or the CaT region. In the next level, we use multiple spectral regions to estimate the corrected velocity of a star from a single spectroscopic observation. Finally, in our hierarchical model, we incorporate multiple spectroscopic observations into our estimate of the corrected velocity of the star.}
		\label{tab:params}

	\end{table*}
    
\begin{figure*}
	\centering
	\includegraphics[width=0.8\textwidth]{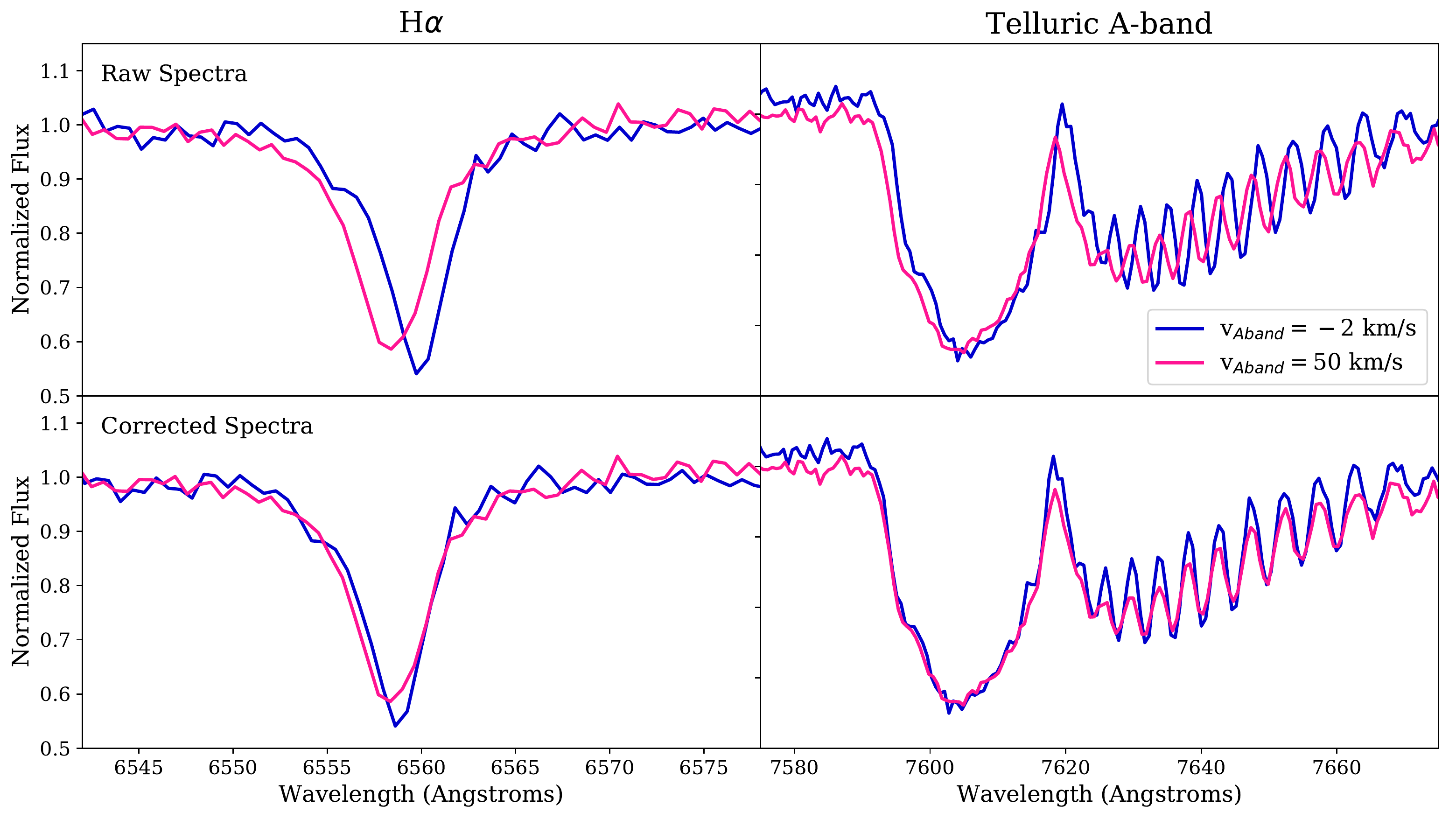}
	\caption{Illustration of the velocity offset caused by slit miscentering. Blue and pink lines show spectra of the same star taken during different observing runs; lefthand panels show the H$\alpha$ region of the spectrum, and righthand panels show the telluric A-band region. Top panels show the raw spectra, in the observed frame, uncorrected for heliocentric motion as well as slit miscentering; the H$\alpha$ lines and telluric absorption lines are clearly misaligned. The lower panels show the spectra with heliocentric and A-band corrections applied.}
	\label{fig:aband}
\end{figure*}

    In this section, we describe in detail the model implemented by the \textsc{Velociraptor} software. As explained in Section \ref{sec:vr_mot}, different observations of the same star will have different raw velocities, due to the motion of the Earth around the Sun as well as slit miscentering. We demonstrate this effect in Figure \ref{fig:aband}, which shows the $H\alpha$ region and the telluric A-band region for two spectra of a relatively bright HALO7D target ($m_{F606W}=19.1$) observed on different nights. The raw spectra are clearly not at the same velocity: while $\sim 5$ km/s of this velocity offset is due to the Earth's motion around the Sun, the remaining 50 km/s offset is entirely due to the misalignment of the object in the slit.
    
    As such, applying these corrections prior to co-adding or stacking spectra is essential in order to accurately estimate the velocity of a star. However, because the A-band correction is measured from an absorption feature, if the spectrum is faint and noisy, the estimate of the A-band correction will also be noisy. 
    
    In order to address these challenges, we present the \textsc{Velociraptor} technique. \textsc{Velociraptor} implements a Bayesian hierarchal model, modeling the raw velocities and A-band corrections of all observations of a star simultaneously. Standard practice would be to stack spectra and then measure a velocity, usually using a cross correlation method (e.g., spec1d; \citealt{Newman2013}) or a maximum-likelihood method (such as the Penalized Pixel Fitting method of \citealt{Cappellari2004}). However, stacking before measuring a velocity neglects the A-band corrections (and associated uncertainties) of different observations. Bayesian hierarchical modeling provides a natural, fully probabilistic framework for incorporating all available information in the spectra while properly accounting for uncertainties. 
    
    In Section \ref{sec:def}, we define terminology used to describe our model. We then explain how we model individual spectroscopic observations in Section \ref{sec:single_mode}. Section \ref{sec:hier_mode} describes the hierarchical model employed to measure the velocity of a star from multiple observations. Details of fake data testing, including sample trace and corner plots, can be found in the Appendix.
	
	\subsection{Definitions}
    \label{sec:def}
	
	We begin by explicitly defining some terminology and symbols used in the description of our method.

	\textit{Spectral Regions}: a segment of a spectrum, a few hundred \AA ngstroms in wavelength, centered on an absorption feature. We use the letter $j$ to denote a specific spectral region; for example, $F_j(\lambda)$ denotes the flux as a function of wavelength for spectral region $j$.

	\textit{Pixels}: Each spectral region contains pixels $i=1, ..., I$. The value $I_j$ denotes the total number of pixels in spectral region $j$.

	\textit{$\lambda_i \rightarrow x_{ij}$}: For evaluating polynomials, we rescale the wavelength array of a given spectral region $j$ onto the range [-1,1]:

	\begin{equation}
		x_{ij}(\lambda_{i})= \frac{2(\lambda_{i}-\lambda_{\mathrm{min},j})}{\lambda_{\mathrm{max},j}-\lambda_{\mathrm{min},j}}-1.
		\label{eqn:lam_x}
	\end{equation}

	 So, $x_{ij}$ denotes the value that $\lambda_i$ takes when rescaled onto the range determined by the range of spectral region $j$: $[\lambda_{\mathrm{min},j}, \lambda_{\mathrm{max},j}]$. Using this definition, $x_{0,j} = x(\lambda_{\mathrm{min},j})=-1$, and $x_{I,j} = x(\lambda_{\mathrm{max},j})=1$.
 
	\textit{Observations}: Each spectrum has $k=1, ..., K$ observations. Therefore, each observation has its own $v_{\mathrm{raw}, k}$, $v_{\mathrm{Aband}, k}$, $v_{\mathrm{helio}, k}$.
    
    \textit{Distributions}: We use standard statistical notation to express random variable distributions, which will include normally-distributed $(x \sim \mathrm{N}(\mu,\sigma^2))$ and Gamma-distributed $(x \sim \mathrm{Gamma}(a,b))$.
    
	\subsection{Single-Mode: Modeling a Single Spectrum}
    \label{sec:single_mode}

	We first present our Bayesian method of estimating the velocity of a star from a single spectrum. To estimate a stellar velocity, we use the spectral regions that contain the most velocity information. We typically use 3 regions: the region around H$\alpha$ ($6500-6650$ \AA) and the Calcium triplet region (CaT; $8450-8700$ \AA) to estimate the raw velocity, and the telluric A-band region ($7500-7750$ \AA) for the A-band correction.  

	To model a region of the spectrum, we first begin with a template. Our templates consist of bright velocity standards that were observed with a very similar configuration to our science spectra. The templates used in this analysis are described in detail by \cite{Toloba2016}; they have high signal-to-noise ratios (100--800 \AA$^{-1}$), and span a range of spectral types (from B1 to M8) and luminosity classes (from dwarfs to supergiants). 
	
	For the HALO7D targets, we use the template HD105546. While this star is a horizontal branch star, its color is consistent with the color range of our targets, and its spectrum has absorption in H$\alpha$ and CaT. In order to estimate the raw velocity of our template star, we use a simple model of a polynomial with inverted Gaussians for the absorption lines. Because the template star was trailed through the slit during observation, it does not suffer from slit miscentering, so its A-band correction is 0 km/s. We verify that no additional correction to the wavelength solution is required by checking the consistency of the velocities measured at H$\alpha$ and CaT.
    
    We use the spectrum of HD105546, shifted to the rest frame, to estimate the velocity of $H\alpha$ and CaT regions of the HALO7D target. We use the same spectrum in the observed frame (i.e., unshifted) to estimate the A-band velocities of the HALO7D targets. We model the different regions of the target spectrum separately, while demanding that the velocities at $H\alpha$ and CaT be the same. 

To model a spectral region, we allow the velocity, absorption line strength, and continuum level to vary. Our vector of parameters, which we denote as $\theta_{\rm Line}$, are the velocity $v_{\rm Line}$, the absorption line strength $C$, and the Legendre polynomial coefficients $b_l$ which control the continuum level. Given that we look at narrow spectral regions, we find $l=1$ (i.e., a straight line with varying slope and intercept) sufficient to model the continuum. 

As a function of scaled wavelength $x$, our model $M(x, \theta_{\rm Line}$) can be written as:

\begin{equation}
	M_{\rm Line}(x, \theta_{\rm Line})=\sum_{l}b_{l}P_{l}(x) \times \frac{ T_{v_{\rm Line}}(x)+C}{1+C},
    \label{eqn:model_line}
	\end{equation}

where $P_l$ are the Legendre polynomials and $T_{v_{\rm Line}}(x)$ is the template flux, shifted to velocity $v_{\rm Line}$.  
    
    The likelihood of the observed spectral flux $F_{\rm Line}$ given the model parameters is thus:

	\begin{equation}
		p(F_{\rm Line}|\theta_{\rm Line}) = \prod_{i=0}^{I} \mathrm{N}(F_{\rm Line}(x_{i})|M_{\rm Line}(x_{i}, \theta_{\rm Line}), \sigma^{2}_{i}),
	\end{equation}
where, for $i=1, ..., I$ pixels, $x_i$ is the rescaled wavelength value, $F_{\rm Line}(x_i)$ is the flux at that rescaled wavelength, $M(x_i,\theta_{\rm Line})$ is the model flux, and $\sigma_i$ is the noise in that pixel (as returned by the spec2d pipeline).

We can write down the posterior probability distribution for our model parameter making use of Bayes' Theorem. Bayes' Theorem gives the probability of a vector of model parameters $\theta$ given a vector of of data $y$:

	\begin{equation}
	p(\theta | y) = \frac{p(y| \theta) p(\theta)}{p(y)},	
    \label{eqn:bayes}
	\end{equation}
where $p(y| \theta)$ is the likelihood of the data given the parameters; $p(\theta)$ is the prior probability of the parameters; and $p(y)$ is the probability of the data (in practice, this term serves as a normalization). In order to sample from the posterior distribution for our model parameters $\theta_{\rm Line}$, we must specify their prior distributions: $p(\theta_{\rm Line} | F_{\rm Line}) \propto p(F_{\rm Line}|\theta_{\rm Line}) p(\theta_{\rm Line})$. 

Our prior distributions are listed in Table \ref{tab:params}; we generally assume reference (i.e., Jeffreys) priors. For the absorption line parameters $C$, we assign a Gamma distribution prior to the quantity $C+1$. The Gamma distribution is defined over the range $x>0$, and is a common prior choice for scale parameters. Because of our chosen parameterization, if $C<-1$ the line becomes an emission line. In addition, as $C$ becomes large, the absorption line becomes indistinguishable from the continuum. We therefore assign a Gamma(2, 2) prior to $C+1$ in order to constrain the possible allowed values for $C$. 

	The total posterior probability of the full vector of spectrum parameters $\theta_{\rm spec}$ is given by the product of the posterior probabilities of the different lines used:

	\begin{equation}
	p(\theta_{\rm spec} | \mathcal{F}) \propto \prod_{j=1}^{J} p(F_{\mathrm{Line}, j}|\theta_{\rm Line}) p(\theta_{\rm Line}),
	\label{eqn:single_mode}
	\end{equation}
where $\theta_{spec}=(v_{\rm raw}, v_{\rm Aband}, C_1,..,C_J, b_{l,1},...,b_{l,J})$ is the full vector of parameters describing the spectrum. Here we are denoting $\mathcal{F}=\lbrace F_{\mathrm{Line},1}, ..., F_{\mathrm{Line}, J}\rbrace$ as the set of fluxes over all spectral regions. When modeling three spectral regions, $\theta_{\rm spec}$ contains 11 free parameters.
    
    In order to sample from the posterior, we use \verb+emcee+ (\citealt{ForemanMackey2013}), a \textsc{python} implementation of the \cite{Goodman2010} affine-invariant Markov chain Monte Carlo (MCMC) ensemble sampler. We first initialize our walkers by estimating the parameters one at a time. Results from extensive fake data testing, including sample trace and corner plots, can be found in the Appendix.

	\subsection{Hierarchical Modeling}
    \label{sec:hier_mode}

	In order to combine spectra from different observations, we employ Bayesian hierarchical modeling. While Bayes' Theorem (Equation \ref{eqn:bayes}) gives the probability of a vector of model parameters $\theta$ given a vector of of data $y$, it is often desirable for the parameters themselves to be drawn from a distribution, whose values we would like to estimate. These \textit{hyperparameters} ($\varphi$), are incorporated into Bayes' Theorem as follows:

	\begin{equation}
	p(\theta, \varphi | y) \propto p(y | \theta,\varphi) p(\theta| \varphi) p(\varphi) .
	\end{equation}

	$p(\theta | \varphi)$ is the probability of the hyperparameters given parameters $\theta$ and $p(\varphi)$ is the hyperprior.

	For our model for multiple observations of a star, we have two hyperparameters: $v$, which is the ``true" velocity of the star, and $\sigma_v$, the dispersion of velocity measurements (this term serves to model additional uncertainty/noise not captured by the reduction pipeline). For the velocity of a star with $K$ observations with spectra $\mathcal{F}_1,...,\mathcal{F}_K$, the full posterior is given by

	\begin{multline}
	p(v, \sigma_v, \theta_{\mathrm{spec}, 1}, ..., \theta_{\mathrm{spec}, K}|\mathcal{F}_1,...,\mathcal{F}_K) =\\
    p(v_{1}, ..., v_{K}|v, \sigma_v) 
    \times \prod_{k=1}^K p(\theta_{\mathrm{spec}, k} | \mathcal{F}_{k}) p(v, \sigma_v)  ,
	\end{multline}
	where $v_k=v_{\mathrm{raw},k}-v_{\mathrm{Aband},k}-v_{\mathrm{helio},k}$ is the corrected velocity for observation $k$, and $p(v, \sigma_v)$ is the prior distribution on the hyperparameters. 
    
    For these measurements, we can consider the fact that we have substantial prior information about the extent to which these velocities should agree: we know that we are observing the same star with each observation. Therefore, it does not make sense for $\sigma_v$ to be arbitrarily large, and a standard non-informative prior is not necessarily the best choice. We assign $\sigma_v^2$ to be drawn from an Inverse-gamma distribution with parameters $a=7, b=72$. This prior distribution has a mean of $b/(a-1)=12$, a mode $b/(a+1)=8$, and variance $b^2/((a-1)^2(a-2))=28.8$. This prior distribution therefore assigns highest probability to $\sigma_v$ in the range of $3-4$ km/s, but does allow for $\sigma_v$ to take on higher values if demanded by the data. 
    
	Given the complexity of our model, we use \verb+emcee+ to sample from the posterior. All our model parameters and prior distributions, for each level of the model, are listed in Table \ref{tab:params}.
	
\begin{figure}
		\centering
		\includegraphics[width=0.35\textwidth]{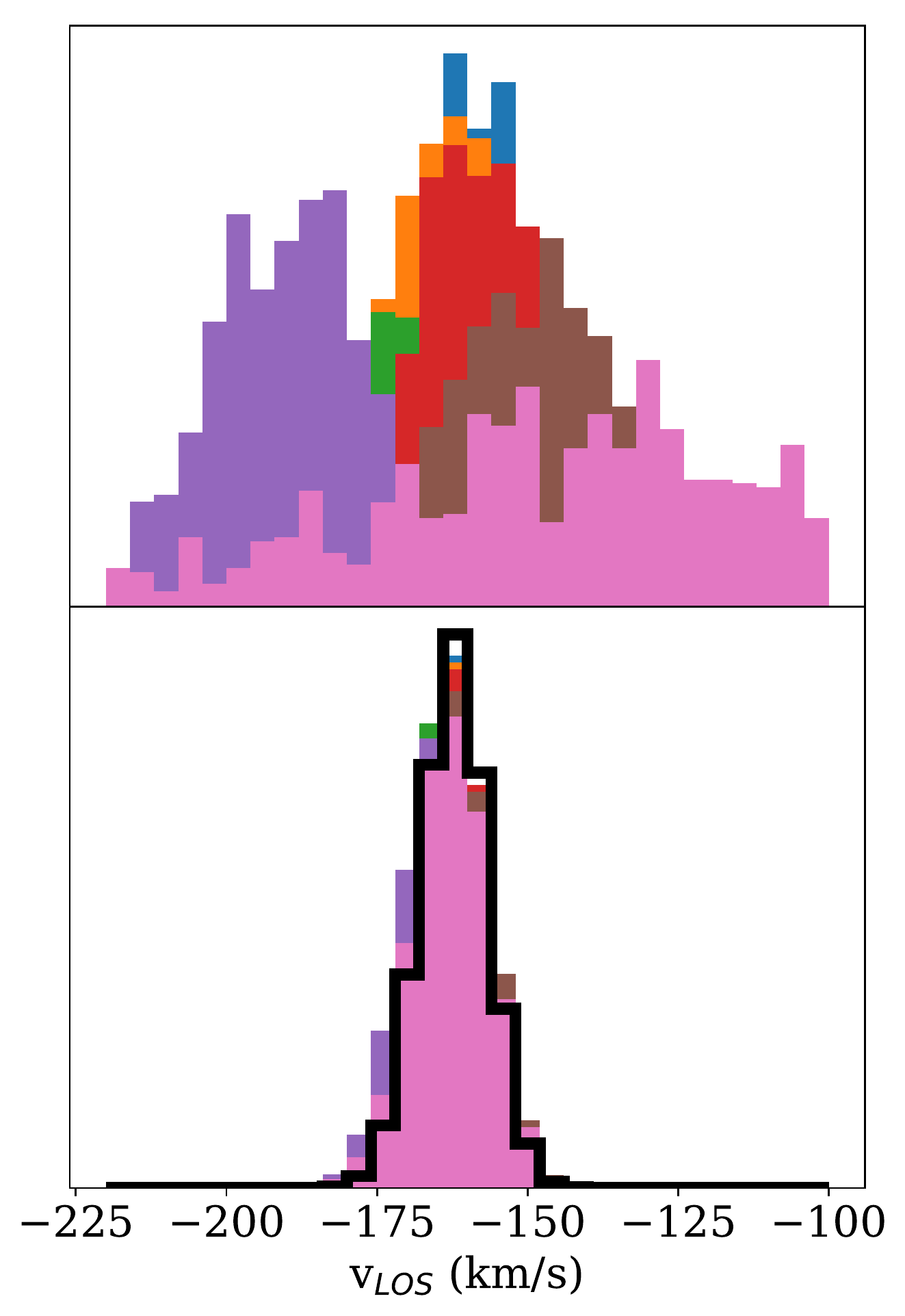}
		\caption{Histograms of posterior samples for the corrected velocity of a HALO7D target ($m_{F606W}=22.0$) from seven observations. The top panel shows the posterior samples for the velocities when the spectra are modeled independently: note that these are the PDFs for $v=v_{raw}-v_{Aband}-v_{helio}$. The bottom panel shows the PDFs for the individual observations once they have been combined into the hierarchical model. The final PDF for the corrected velocity (thick black line), incorporating all observations, thus folds in all information and uncertainty from all observations of a star.}
		\label{fig:vr_pdfs}
	\end{figure}

	A demonstration of this technique is shown in Figure \ref{fig:vr_pdfs}, for a HALO7D target with $m_{F606W}=22.0$. This target was observed seven times, in varying conditions, over the course of Spring 2015. Each of the histograms in the top panel represents the posterior distribution for the \textit{corrected} velocity for each of these seven observations. The varying widths of these PDFs reflect the varying quality in observing conditions across the different nights of observing. These posterior samples were derived using \verb+emcee+ to sample the posterior distribution given in Equation \ref{eqn:single_mode}. However, once we link the observations by combining them with the hierarchical model, the posterior distributions for the individual velocities converge (lower panel). The posterior distribution for the corrected velocity of the star is shown in black: this posterior incorporates all information, as well as sources of uncertainty, from the seven observations.
	
\begin{figure}
		\centering
		\includegraphics[width=0.45\textwidth]{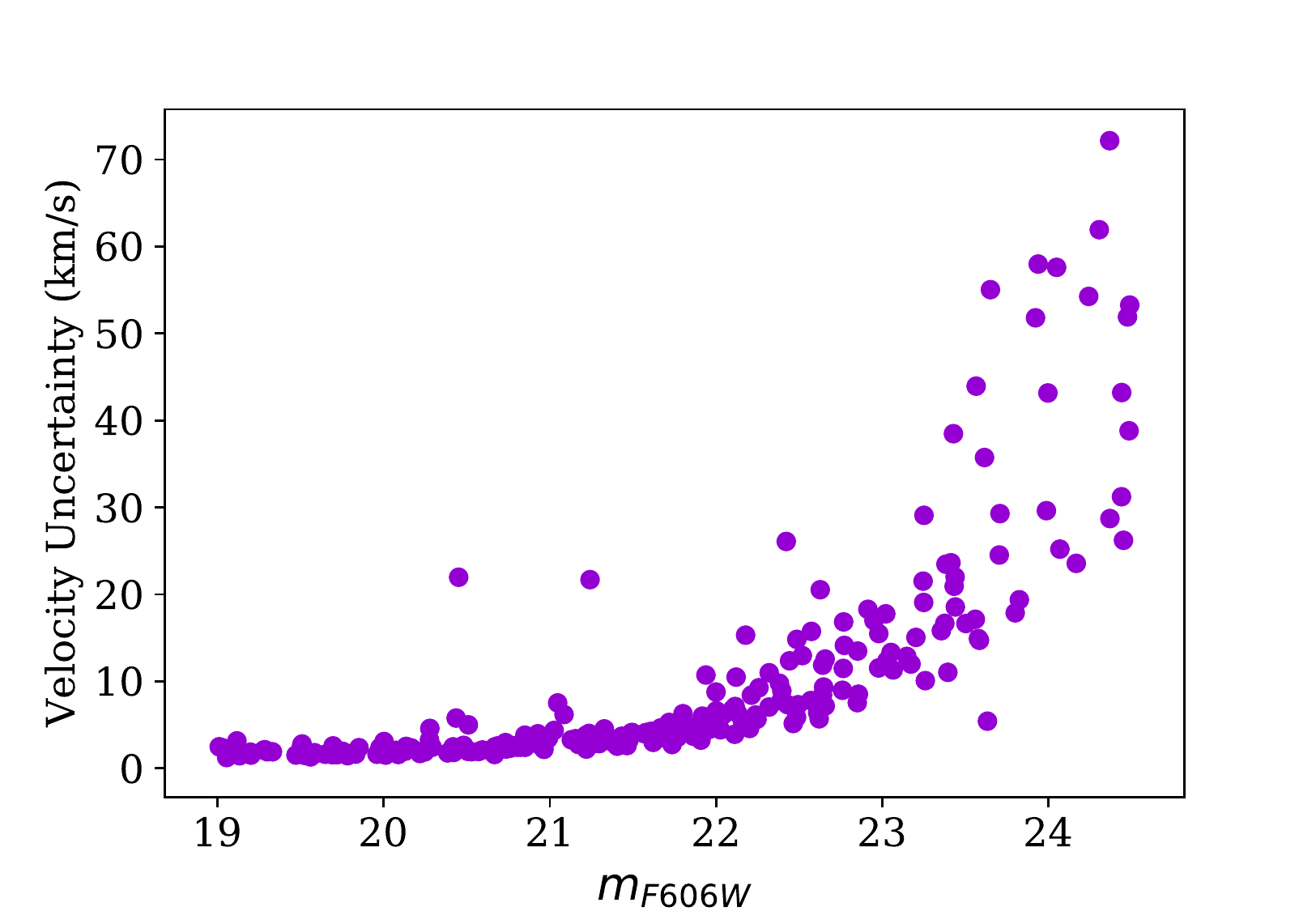}
		\caption{Velocity uncertainties for the HALO7D sample as a function of $m_{F606W}$ apparent magnitude. Velocity uncertainties are the 16th and 84th percentiles of the posterior distributions.}
		\label{fig:vr_err}
	\end{figure}

	The resulting velocity uncertainties for the HALO7D targets as a function of $m_{F606W}$ apparent magnitude are plotted in Figure \ref{fig:vr_err}. Velocity errors are computed as half the difference between the 84th and 16th posterior percentiles: $v_{err}=(v_{84}-v_{16})/2$. At the bright end of our sample, our velocity uncertainties are as low as 1-2 km/s; velocity uncertainties remain below 10 km/s for stars brighter than $m_{F606W}=22.$ Stars at the faint end of our sample reach velocity uncertainties as high as $\sim 50$ km/s.
	
	For more details on testing our method on fake data, including sample trace and corner plots, please see the Appendix. 
	\section{Results}
	\label{sec:results}
	
	\subsection{LOS Velocity Distributions}
	\label{subsec:los_dist}
	
	We use the velocities measured from \textsc{Velociraptor} to study the LOS velocity distributions of the stellar halo. Heliocentric LOS are converted to the Galactocentric Standard of Rest (GSR) frame by assuming a circular speed of 240 km s$^{-1}$ at the position of the Sun ($R_0=8.5$ kpc), with solar peculiar motion $(U,V,W)=(11.1, 12.24, 7.25)$ km s$^{-1}$ (\citealt{Schonrich2010}). 
	
\begin{figure*}
	\includegraphics[width=0.5\textwidth]{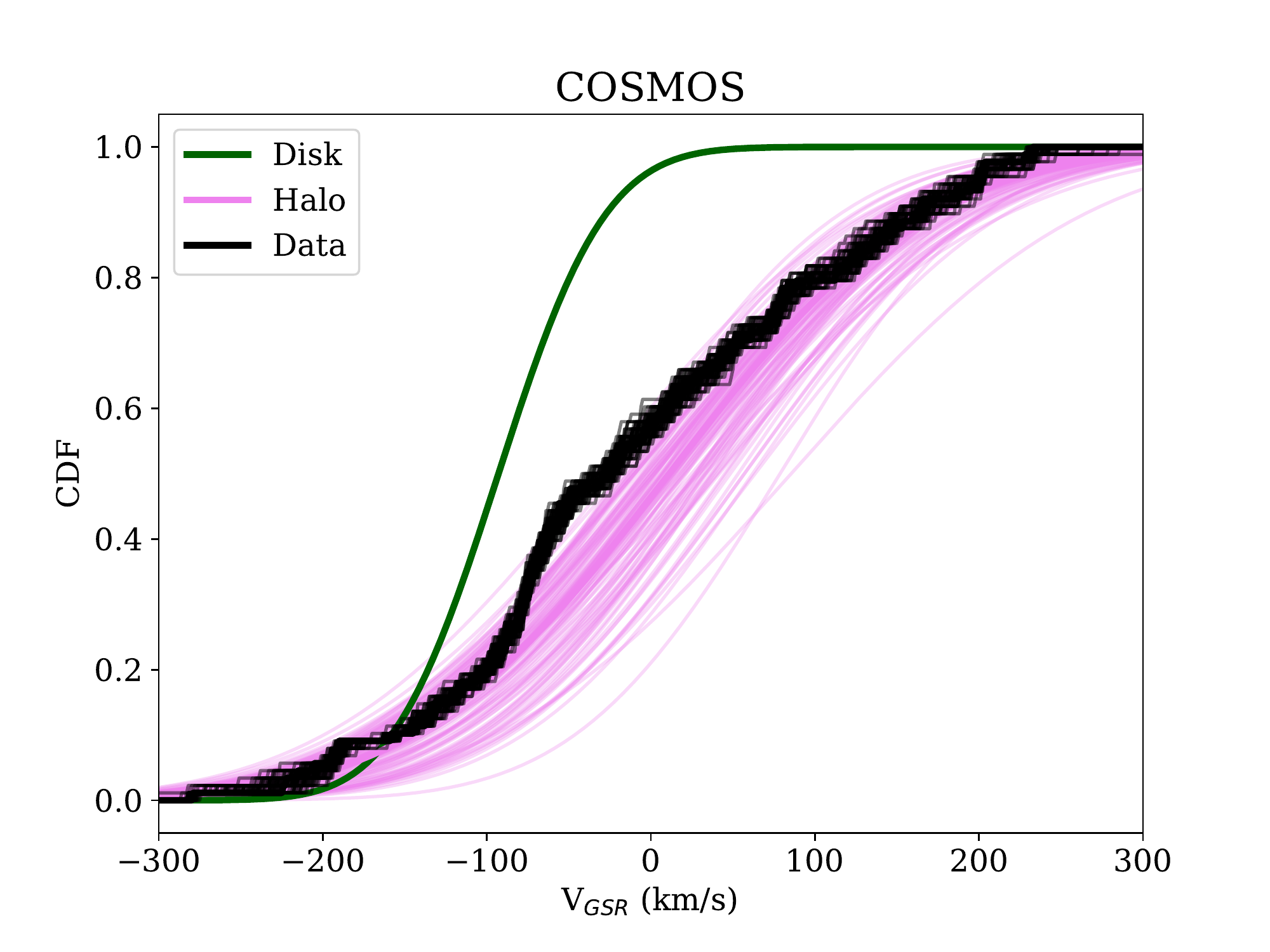}
	\includegraphics[width=0.5\textwidth]{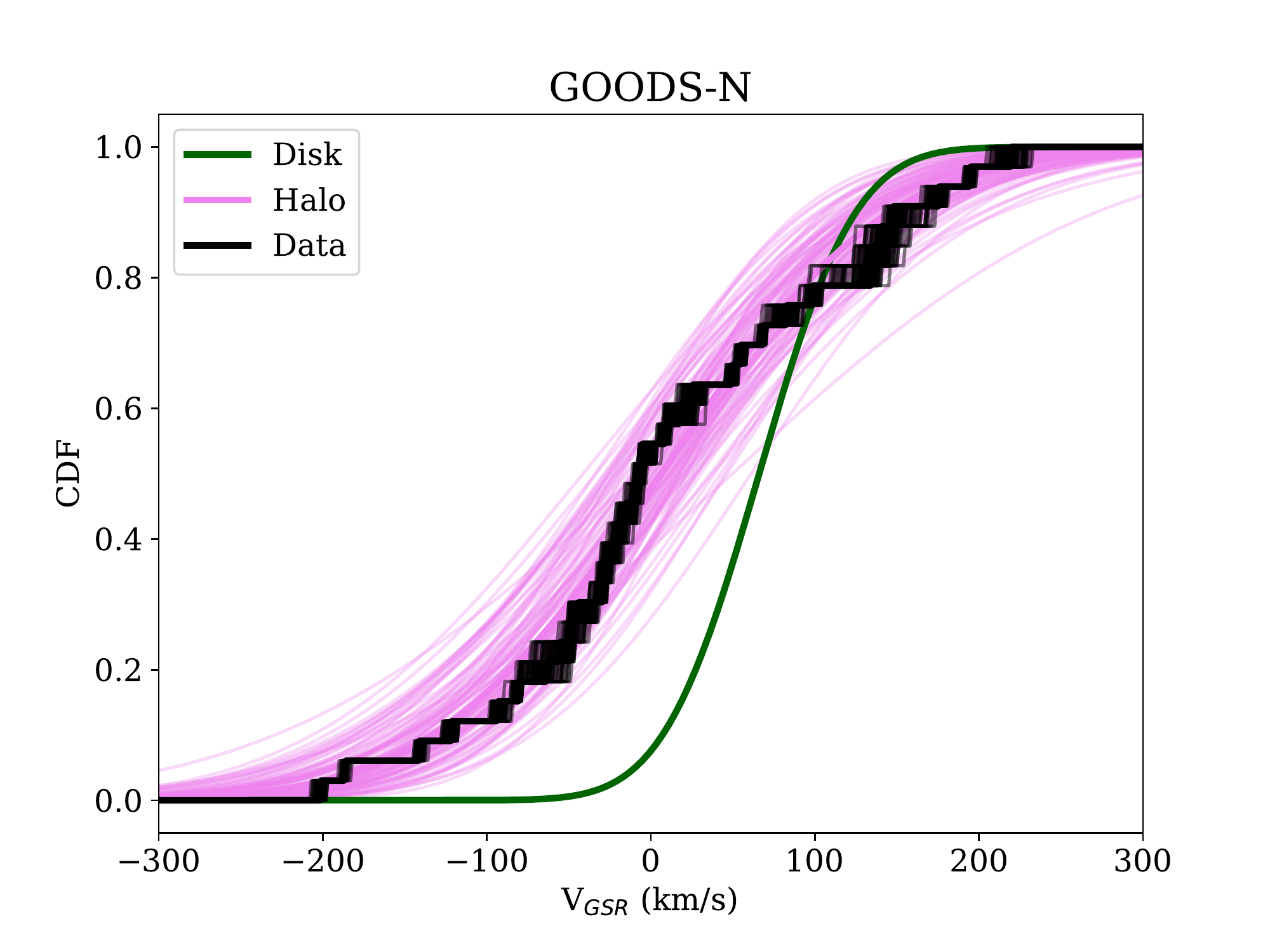}
	\includegraphics[width=0.5\textwidth]{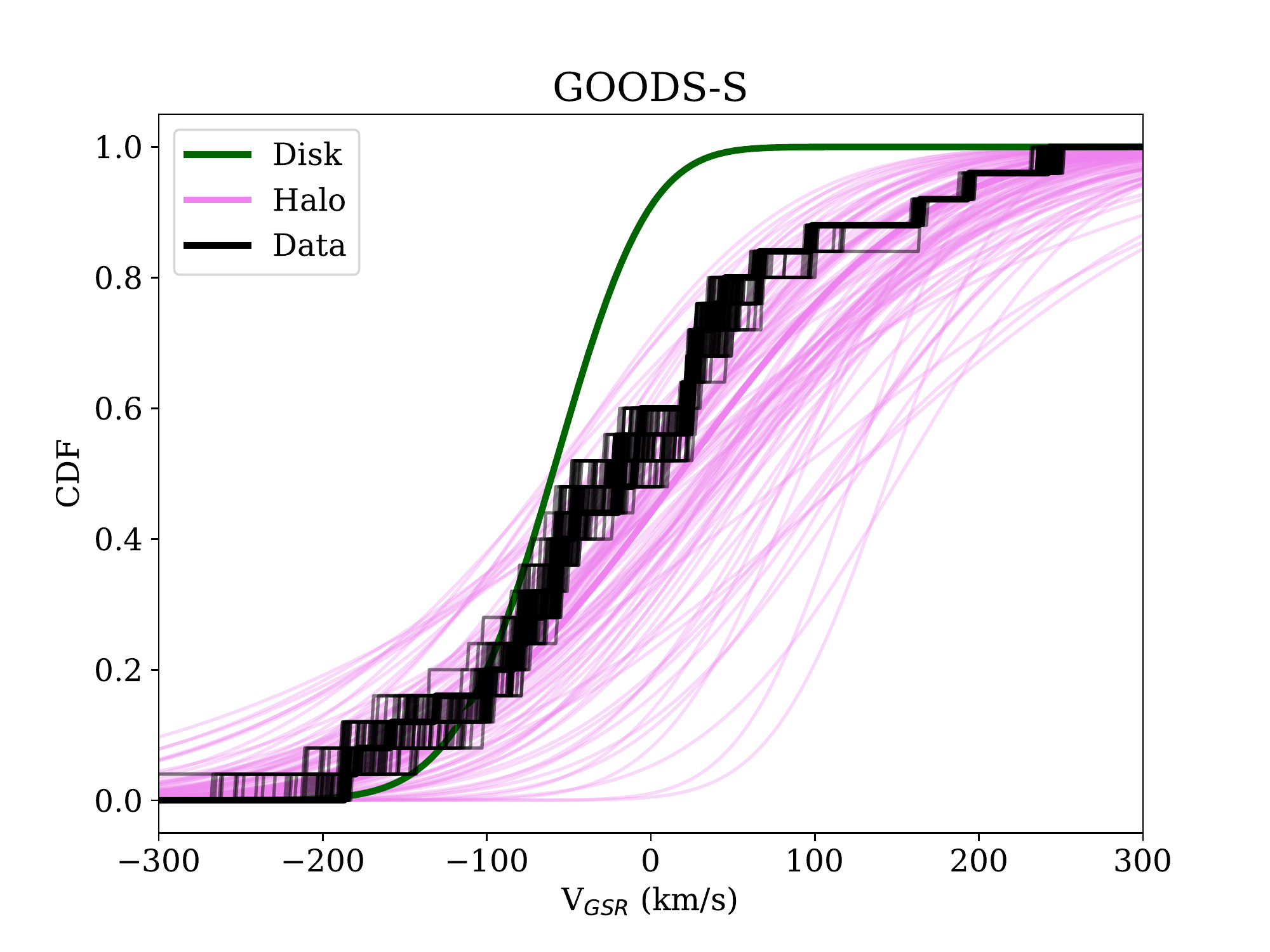}
	\includegraphics[width=0.5\textwidth]{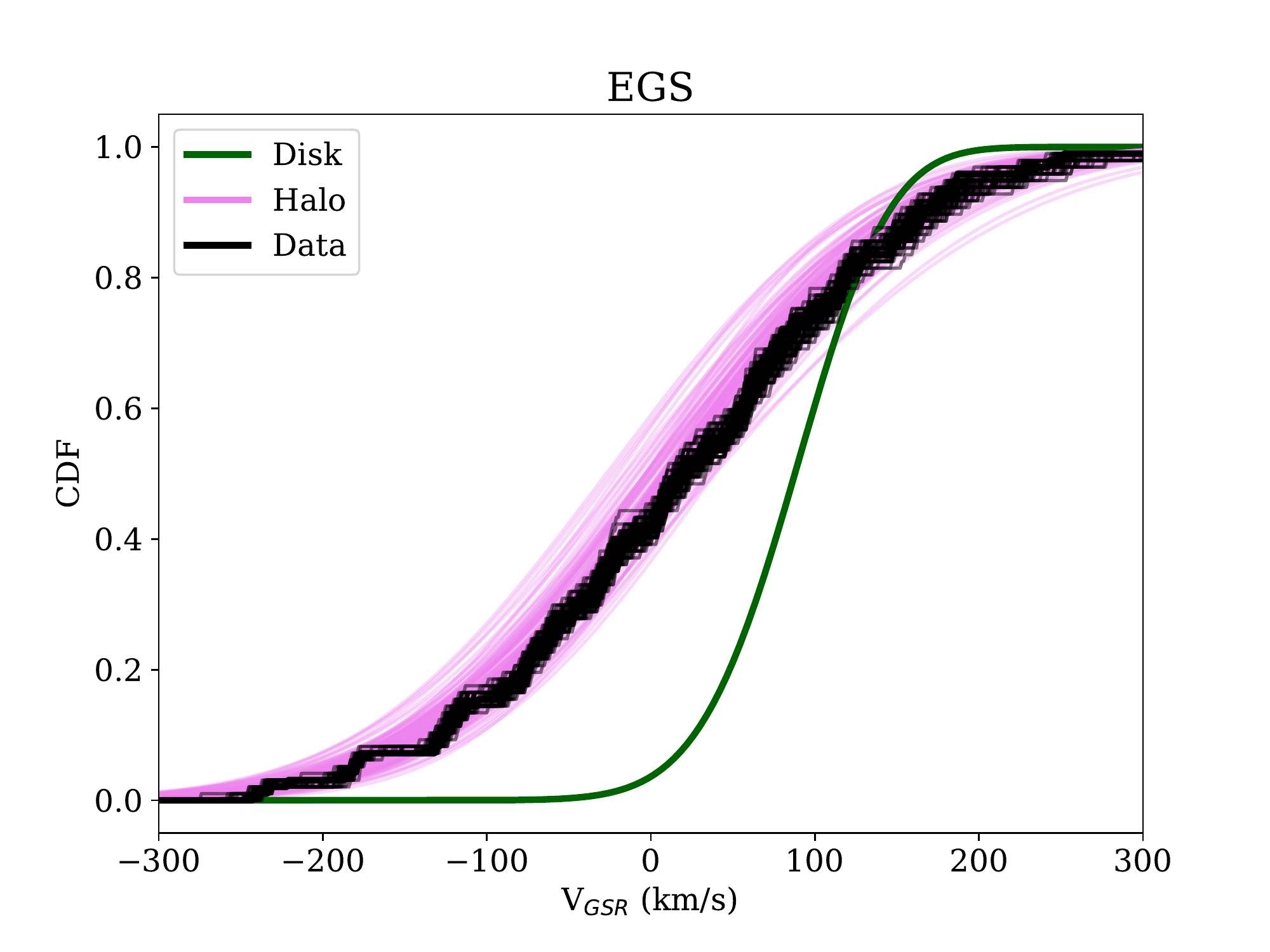}
	\caption{Cumulative LOS velocity histograms in the four HALO7D fields. Velocities are given with respect to the Galactic Standard of Rest (GSR). Black lines indicate the CDFs for the data: for each of the 100 black lines, velocity values were drawn from the posterior distributions for the measurements. The width of each step thus demonstrates the velocity uncertainty for that data point. Pink lines show the halo CDF for 100 draws from the posterior distribution for $\sigma_{LOS}$. The green lines indicate the CDF for the disk model.}
	\label{fig:los_cdf}
\end{figure*}

\begin{figure*}
	\centering
	\includegraphics[width=0.45\textwidth]{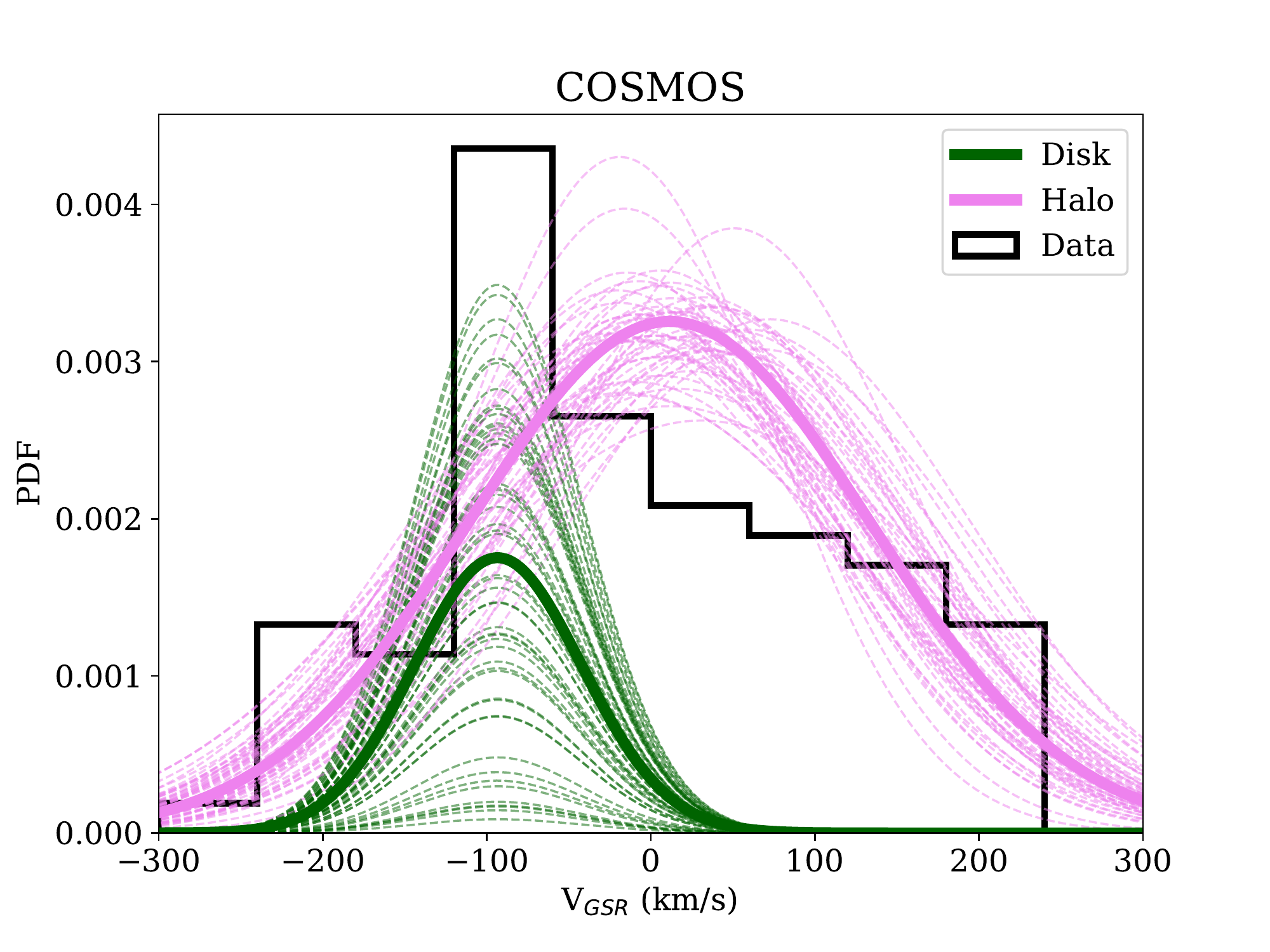}
	\includegraphics[width=0.45\textwidth]{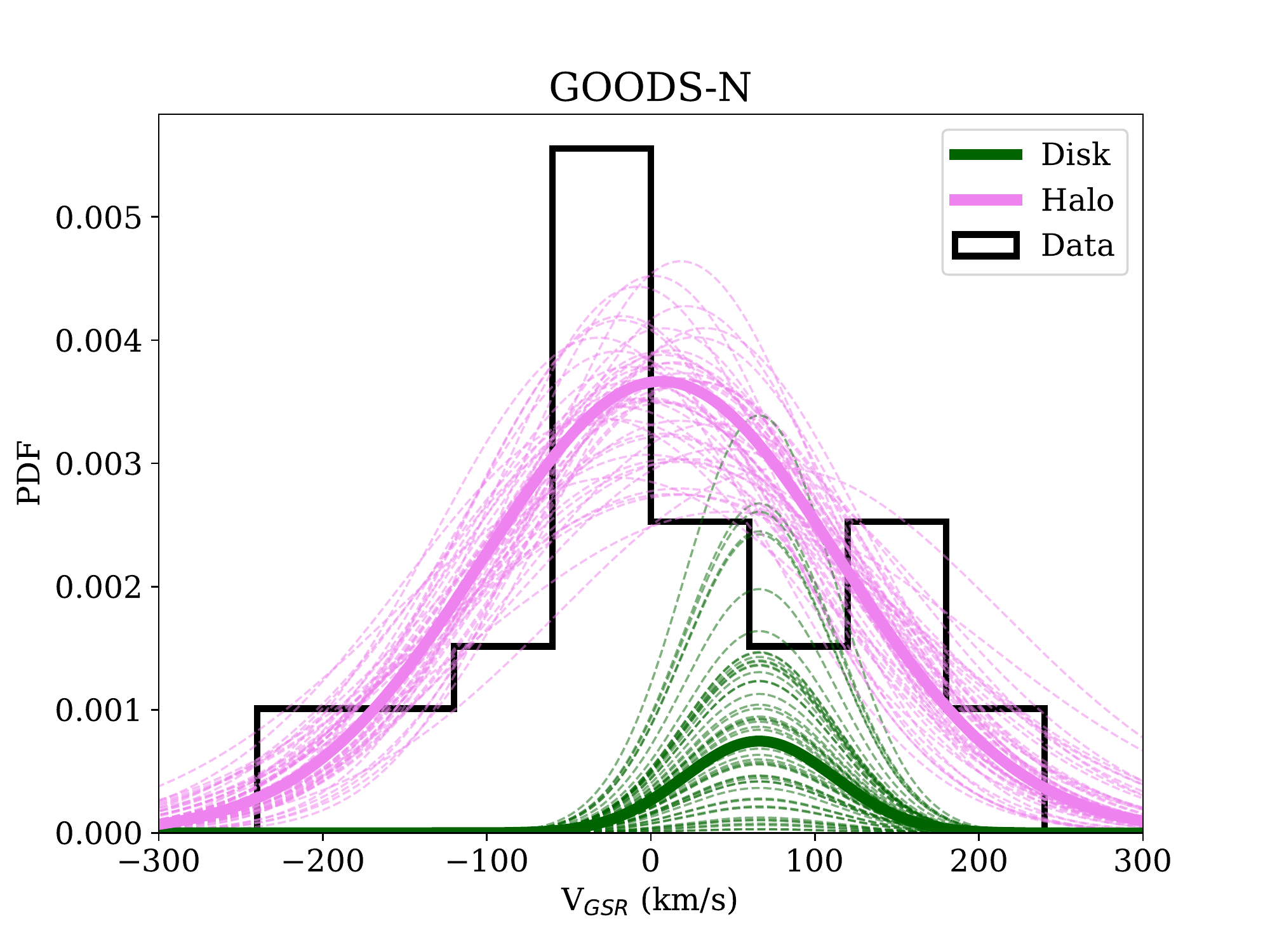}
	\includegraphics[width=0.45\textwidth]{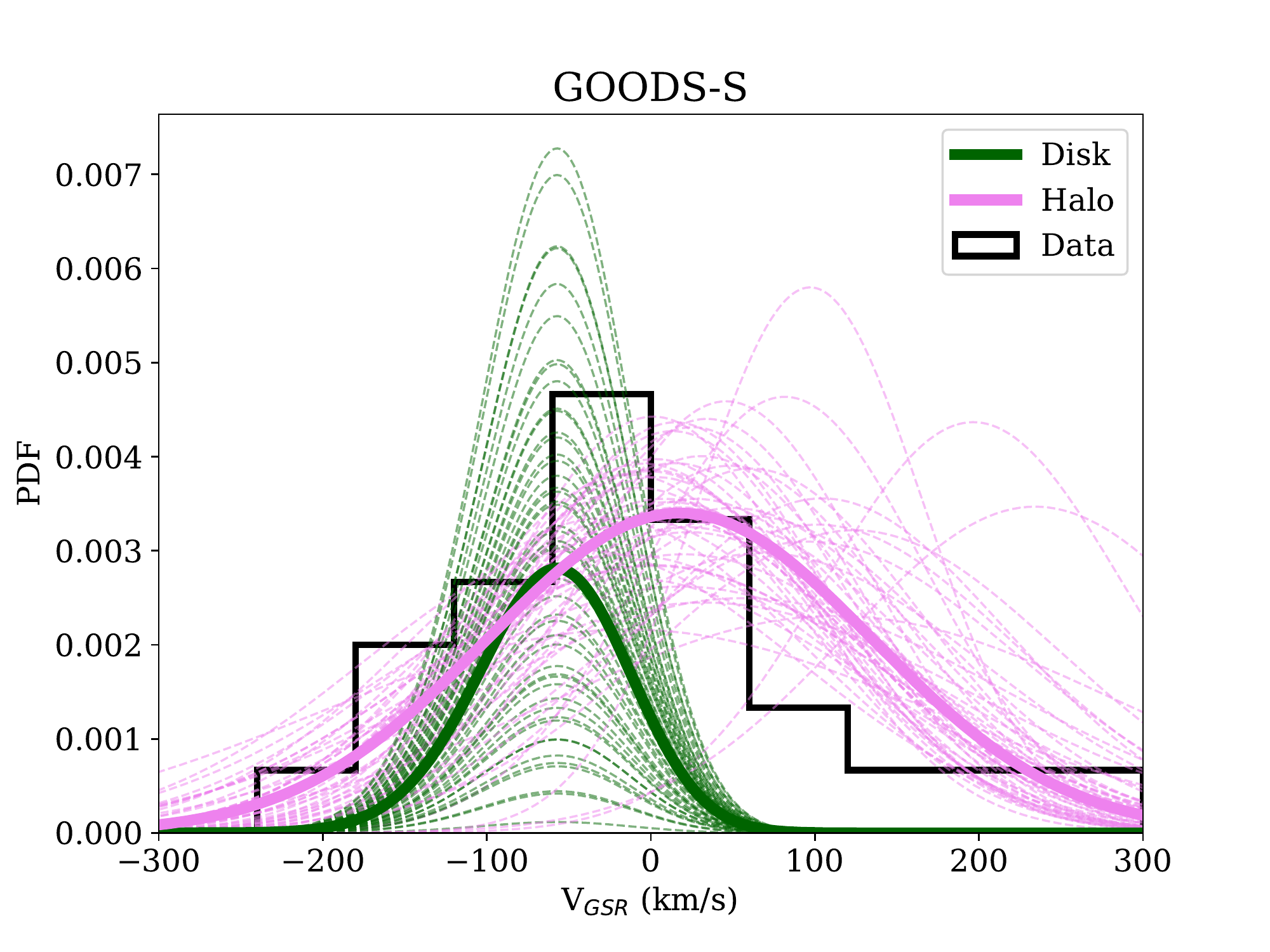}
	\includegraphics[width=0.45\textwidth]{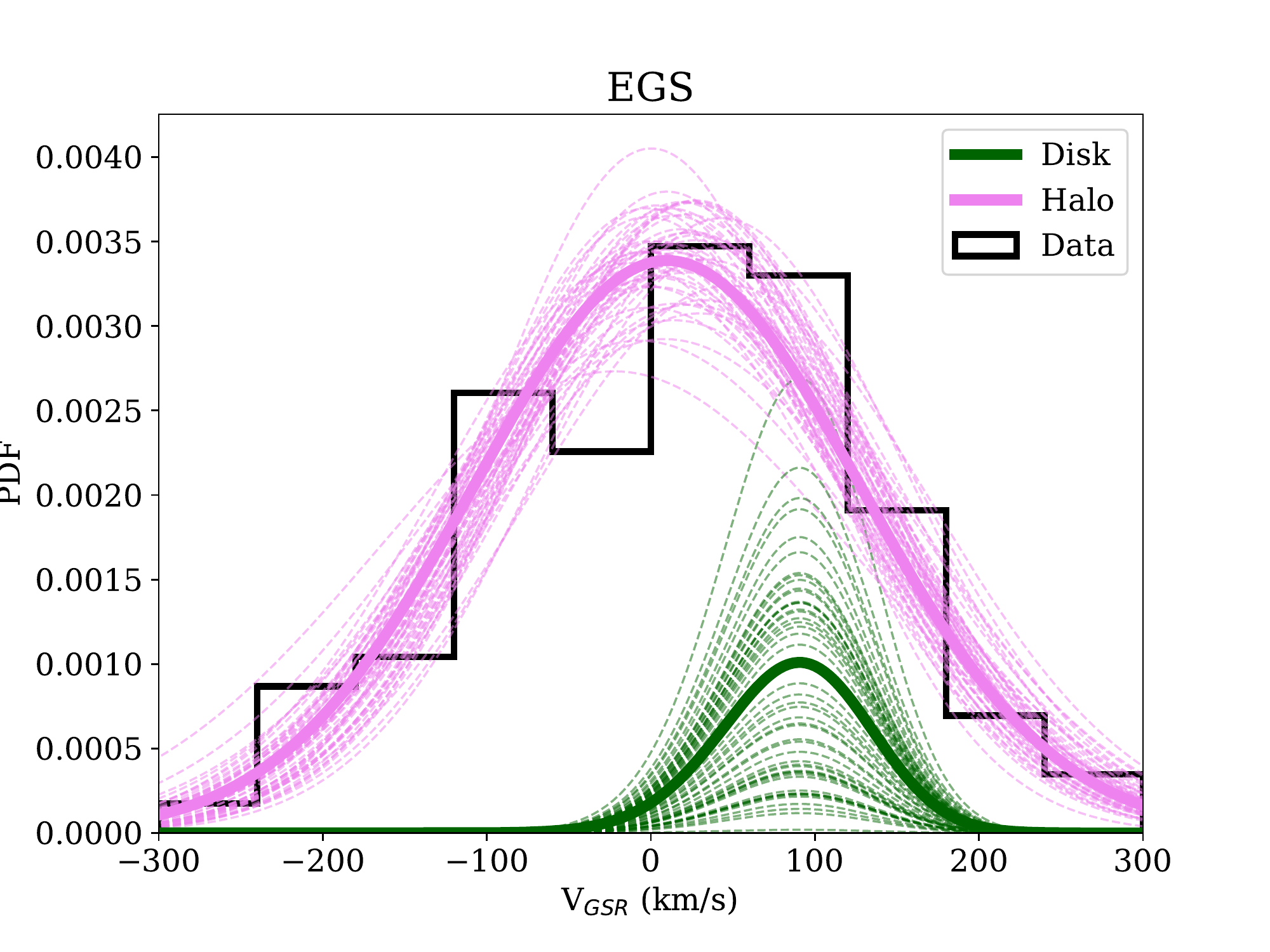}
	\caption{LOS velocity histograms in the four HALO7D fields. Shown in pink are the resulting velocity distributions from 50 draws from the posteriors for $\langle v_{LOS} \rangle$ and $\sigma_{LOS}$. The green line indicates the disk distribution. The parameters of the disk velocity distribution are fixed; only the fraction to the total contribution is allowed to vary. Bold lines show the corresponding distributions for the median posterior values of $f_{\rm disk}$, $\langle v_{LOS} \rangle$ and  $\sigma_{LOS}$.}
	\label{fig:los_hist}
\end{figure*}

	Figure \ref{fig:los_cdf} shows cumulative histograms for the LOS velocity distributions (in the GSR frame) for the four HALO7D fields. To capture the effects of our velocity uncertainties, we have plotted 100 realizations of the velocity cumulative distribution, each time drawing a new value for every velocity from its posterior. Therefore, the apparent thickness of a given step in the histogram is an indication of the uncertainty of that measurement. We also show traditional histograms of the LOS velocities in Figure \ref{fig:los_hist}.
    
Based on the histograms, we see that our samples across all four fields are dominated by a ``hot halo'' population; while their could be hints of substructure in these fields, we find that none of our fields are dominated by kinematically cold substructure, which would appear as narrow ($5-15$ km/s) peaks in the velocity distributions. We leave the discussion of the search and characterization of potential substructure in these fields to future work, where we will also utilize PMs and abundances.
	
	To estimate the LOS velocity dispersion of the halo, $\sigma_{LOS}$, we model the LOS velocity distributions as a two component mixture model, with a halo component and a disk component. We model the halo velocity distribution as a normal distribution with unknown mean and variance: $v\sim\mathrm{N}(\langle v_{LOS} \rangle, \sigma_{LOS}^2)$. 
    
    We model the disk velocity distribution along each line of sight as a skew-normal distribution, with skew parameter $\alpha$, location parameter $\zeta$, and scale parameter $\omega$. The likelihood of an observed velocity $v_i$ given disk parameters is given by:
    
    \begin{equation}
     p(v_i|\alpha, \zeta, \omega)=\frac{2}{\omega}\phi \left( \frac{v_i-\zeta}{\omega}\right)\Phi\left(\alpha \left(\frac{v_i-\zeta}{\omega}\right)\right),
    \end{equation}
where $\phi(x)$ is the standard normal PDF and $\Phi(x)$ is the standard normal CDF. We fix the parameters of the disk velocity distribution, but leave the fraction of disk contamination as a free parameter. We denote our disk PDF as $g_{\rm disk}(v_i)=p(v_i|\alpha, \zeta, \omega)$.

To determine the parameters of our disk model, we use the Besan\c{c}on Galaxy Model (\citealt{Robin2003}). We use synthetic catalogs of 1 square degree areas centered on the coordinates of our survey fields (the larger area gives us better statistics for our simulated disk and halo populations). We then determine the velocity distribution of the (non-WD) disk contaminants within the HALO7D selection boxes, and fit a skew normal to this distribution. The resulting parameters for the disk distributions can be found in Table \ref{tab:results}; they are also plotted in Figures \ref{fig:los_cdf} and \ref{fig:los_hist}.
	
	For this mixture model of disk and halo, the likelihood of a given LOS velocity observation $v_i$, with error $\sigma_{v,i}$, is given by
    
    \begin{equation}
    \begin{split}
    p(v_i|f_{\rm disk},\langle v_{LOS} \rangle, \sigma_{LOS}, g_{\rm disk}) = f_{\rm disk}\times g_{\rm disk}(v_i)\\
   + (1-f_{\rm disk}) \times N(v_i|\langle v_{LOS} \rangle, \sigma_{LOS}^2+\sigma_{v,i}^2). \\
    \end{split}
    \end{equation}
	
	In order to sample from the posterior distribution for our model parameters, we first must assign prior distributions. We assign a standard uniform $[0,1]$ prior on the fraction of disk contamination, and we assign the Jeffreys prior to the mean and dispersion for the halo LOS velocity distribution ($p(\langle v_{LOS} \rangle,\sigma_{LOS}) \propto 1/\sigma_{LOS}$).

	Our posterior is thus:
	
	\begin{equation}
    \begin{split}
		p(\langle v_{LOS} \rangle,\sigma_{LOS}, f_{\rm disk}|v) \propto p(\langle v_{LOS} \rangle,\sigma_{LOS}) p(f_{\rm disk}) \\
      \times \prod_{i=1}^{N_{stars}} p(v_i|\langle v_{LOS} \rangle,\sigma_{LOS}, f_{\rm disk}).
      \end{split}
	\end{equation} 
	
	We use \verb+emcee+ to sample from this posterior. We used 500 walkers, ran the sampler for 1000 steps, and discarded the first 800 steps as burn-in. Median posterior values, along with error bars from the 16/84 percentiles, are quoted for the three model parameters in Table \ref{tab:results}.
	
    Posterior draws are overplotted on the histograms in Figures \ref{fig:los_cdf} and \ref{fig:los_hist}. Each pink line in Figure \ref{fig:los_cdf} is the CDF corresponding to a draw from the posterior for $\langle v_{LOS} \rangle$ and $\sigma_{LOS}$. In Figure \ref{fig:los_hist}, the amplitude of the disk PDFs reflects the uncertainty in the disk contribution. Thicker pink and green lines indicate the distributions corresponding to the median posterior values.
    
\begin{figure*}
	\includegraphics[width=\textwidth]{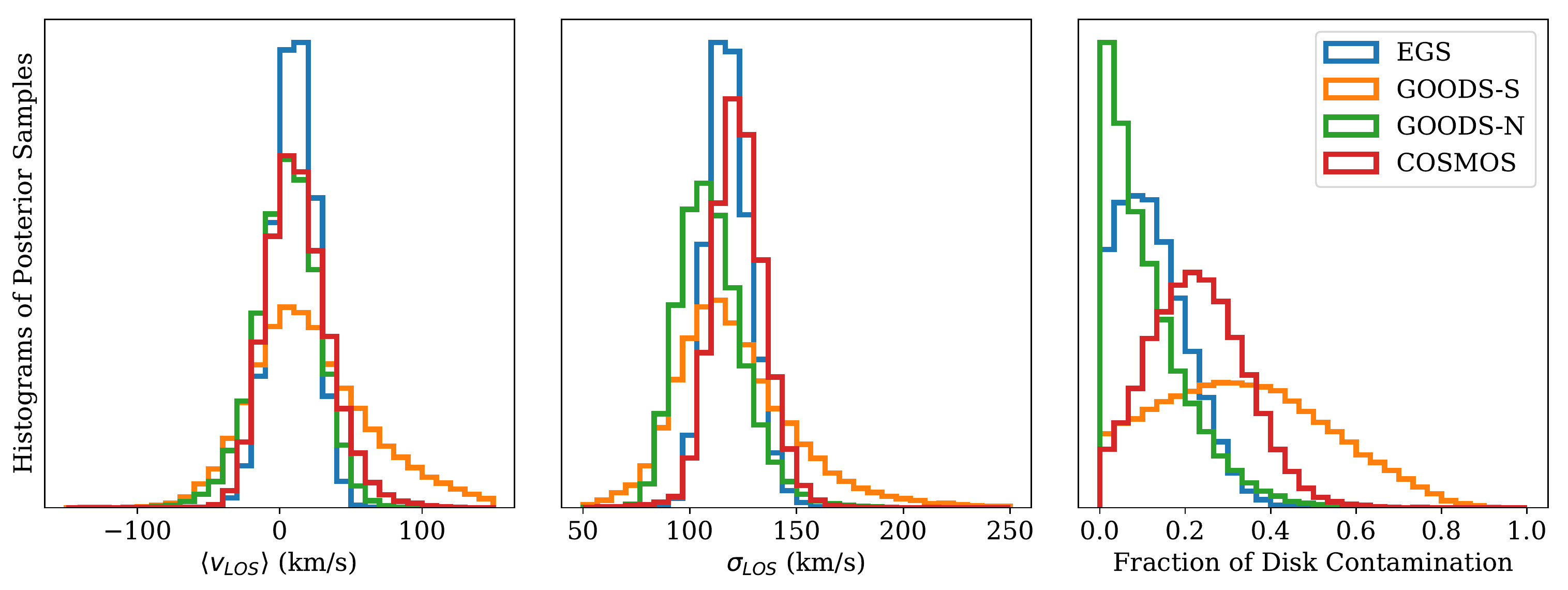}
	\caption{Posterior samples for the mean LOS velocity (right), LOS velocity dispersion (middle) and the fraction of disk contamination (right) for all four fields. All fields have mean LOS velocity consistent with 0 km/s, and dispersions consistent with one another.}
	\label{fig:post_hist}
\end{figure*}

	Histograms of posterior samples for our three free parameters are shown in Figure \ref{fig:post_hist}. The left panel shows the posterior distributions for $\langle v_{LOS} \rangle$; all fields have mean LOS velocity consistent with $0$ km/s. The middle panel of Figure \ref{fig:post_hist} shows the posterior samples for $\sigma_{LOS}$; posterior PDFs for $\sigma_{LOS}$ are consistent across the four fields. The widths of the individual PDFs vary according to the sample size in a given field, but the PDFs substantially overlap. In the COSMOS field, we estimate $\sigma_{LOS}=123^{+12}_{-11}$ km/s; in GOODS-N, $\sigma_{LOS}=110^{+16}_{-13}$ km/s; for GOODS-S, $\sigma_{LOS}=122^{+30}_{-21}$ km/s; and in EGS, we find $\sigma_{LOS}=118^{+11}_{-9}$ km/s. 

The right panel of Figure \ref{fig:post_hist} shows posterior samples for $f_{disk}$ in the four fields. We note that no color or distance information is incorporated into our estimates of the disk contamination, and that this estimate is based on LOS velocities alone. Our estimates of our disk contamination will be more accurate once PM and photometric information are incorporated. EGS and GOODS-N show $0-10$\% disk contamination, consistent with the predictions from Besan\c{c}on (see Table \ref{tab:results}). COSMOS, our lowest latitude field, shows a slightly higher level of disk contamination ($\sim 25\%$, while predicted to be $\sim 11 \%$). The posterior distribution for disk contamination in GOODS-S, our field with the smallest sample size, is very broad, with a posterior median of $30\%$, much higher than the $7\%$ predicted by Besan\c{c}on. We note that this high posterior median is largely due to the small sample size, and that the disk contamination in GOODS-S is poorly constrained based on LOS velocities alone.

    \begin{table*}
	\begin{center}
	\begin{tabular}{c | c  c c c c c c c}
    \hline
		Field & $\langle v_{LOS} \rangle$ (km/s) & $\sigma_{LOS}$ (km/s) & $f_{\rm disk}$ & Predicted $f_{\rm disk}$ & $\alpha$ & $\zeta$ (km/s) & $\omega$ (km/s) & $\langle D \rangle$ (kpc)\\
		\hline \hline
		COSMOS &$13_{-19}^{+23}$ &$123_{-11}^{+12}$ &$0.23_{-0.11}^{+0.12} $ & 0.11 & $0.6$ & $-115$ & $56$ & 21\\
		GOODS-N &$6_{-21}^{+20}$ & $110_{-13}^{+16}$&$0.11_{-0.06}^{+0.12}$ & 0.07 & $0.0$ & $66$ & $46$ & 20 \\
		GOODS-S & $24_{-35}^{+48}$&$122_{-21}^{+30}$ & $0.34_{-0.20}^{+0.22}$ & 0.07 &$-1.1$&$-27$&$58$ &23 \\
		EGS & $9_{-14}^{+15}$& $118_{-9}^{+11}$& $0.13_{-0.08}^{+0.10}$ & 0.07 &$-1.1$ & $121$& $58$& 21\\
        \hline \hline
	\end{tabular}
	\end{center}
	\caption{Summary of results from the modeling of the LOS velocity distributions. Posterior medians are quoted, with error bars giving the 16/84 percentile credible regions. We list the predicted fraction of disk contamination from Besan\c{c}on; because we have included no color or apparent magnitude information into our estimate of the disk contamination, we expect our $f_{\rm disk}$ estimates are higher than the actual disk contamination in our sample. We list the parameters for the disk model for each field: the skew $\alpha$, the location $\zeta$, and the scale $\omega$. These values were derived by fitting a skew normal distribution to the velocities of disk stars that fall in the HALO7D selection box in the Besan\c{c}on Galaxy Model. Finally, we list the average distance to each field, as computed in Equation \ref{eqn:d_pdf}.}
	\label{tab:results}

\end{table*}

	\subsection{Comparison with Other Tracers}
	
	In this section, we compare our results for the HALO7D LOS velocity distributions with other studies conducted over a similar distance range (though using different tracer populations). In order to compare our measurements of $\sigma_{LOS}$ with other studies, we first need to estimate the distance range probed by our sample. We estimate these distance distributions in a similar method to that of D13 and C16, using weighted isochrones to derive the PDF for the absolute magnitude $M_{F814W}$ of a star given its $m_{F606W}-m_{F814W}$ color.  

\begin{figure}
		\includegraphics[width=0.5\textwidth]{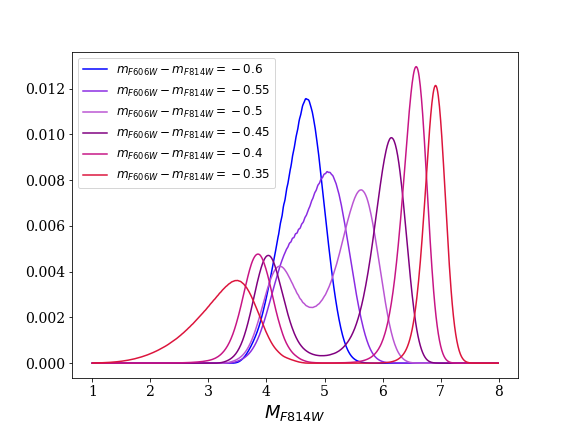}
		\caption{Normalized PDFs for absolute magnitude for six different choices of $m_{F606W}-m_{F814W}$ color. These PDFs are derived from the KDE constructed from the \cite{Vandenberg2006} isochrones, weighted by a Salpeter IMF and the approximate age and metallicity distributions of the halo.}
		\label{fig:color_pdfs}
	\end{figure}

    We weight the \cite{Vandenberg2006} isochrones according to a Salpeter IMF, an age and a metallicity distribution typical of halo stars. We assume that the age and metallicity distributions are Gaussian, with $\langle T \rangle = 12$ Gyr, $\sigma_T=2$ Gyr (e.g., \citealt{Kalirai2012}), $\langle [\mathrm{Fe}/\mathrm{H}] \rangle=-1.9$ and $\sigma_{\mathrm{Fe}/\mathrm{H}} = 0.5$ (e.g., \citealt{Xue2008}). We model the resulting weighted CMD with a kernel density estimate (KDE), using a kernel bandwidth of 0.025. The resulting PDFs for $M_{F814W}$, for six different colors, are shown in Figure \ref{fig:color_pdfs}. 
	
\begin{figure}
		\includegraphics[width=0.5\textwidth]{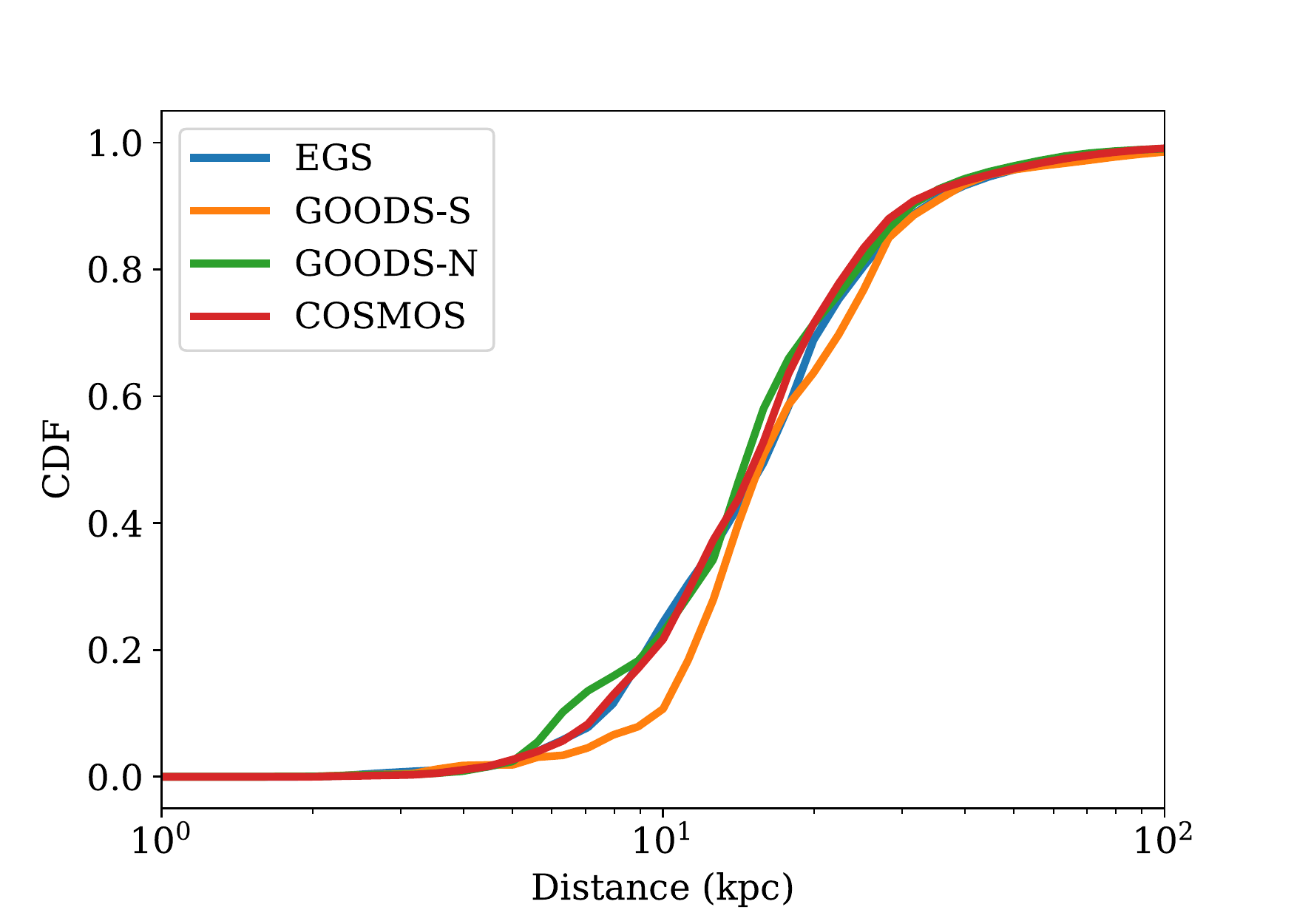}
		\caption{Cumulative distance distributions for the four HALO7D fields. Distance distributions are computed as given by Equation \ref{eqn:d_pdf}, using colors and assuming a MW stellar density profile (\citealt{Deason2011b}). All fields have $\langle D \rangle \sim 20$ kpc.}
		\label{fig:distance_dist}
	\end{figure}

	Using the PDF for absolute magnitude as a function of color in conjunction with the halo density profile (\citealt{Deason2011b}), we derive the PDF for the log distance distribution to our sample:
	
	\begin{multline}
	p(\log D|m_{F814W}, m_{F606W}, \rho) \propto  p(\log D | \rho) \times \\
    \sum_{n=1}^{N_{obj}} p(M_{F814W}(\log D)|m_{F606W,n}-m_{F814W,n}),\\
	\label{eqn:d_pdf}
	\end{multline}

	where $p(\log D | \rho)$ is the probability of $\log D$ given the \cite{Deason2011b} density profile, and $p(M_{F814W}(\log D)|m_{F606W,n}-m_{F814W,n})$ is the probability of object $n$ having absolute magnitude $M_{F814W}(\log D)$ given its color $m_{F606W,n}-m_{F814W,n}$.\footnote{In the COSMOS field, in the area where we used additional catalogs for selection, we used the Subaru B and V photometric measurements from \cite{Capak2007} and converted these to STMAG F606W as directed by \cite{Sirianni2005}. We used the F814W magnitudes as published in the \cite{Leauthaud2007} catalog.} We then estimate the mean distance to each field $\langle D \rangle = \int D \times p(\log D) d\log D$. Each of the four fields has an average distance $\langle D \rangle \sim 20$ kpc. Figure \ref{fig:distance_dist} shows the cumulative logarithmic distance PDFs of our samples across the four fields, and average distances to each field are listed in Table \ref{tab:results}. We note that no kinematic information is incorporated into our distance estimate, and that information from 3D kinematics will improve our distance estimates in subsequent work.
	
\begin{figure}
	\includegraphics[width=0.5\textwidth]{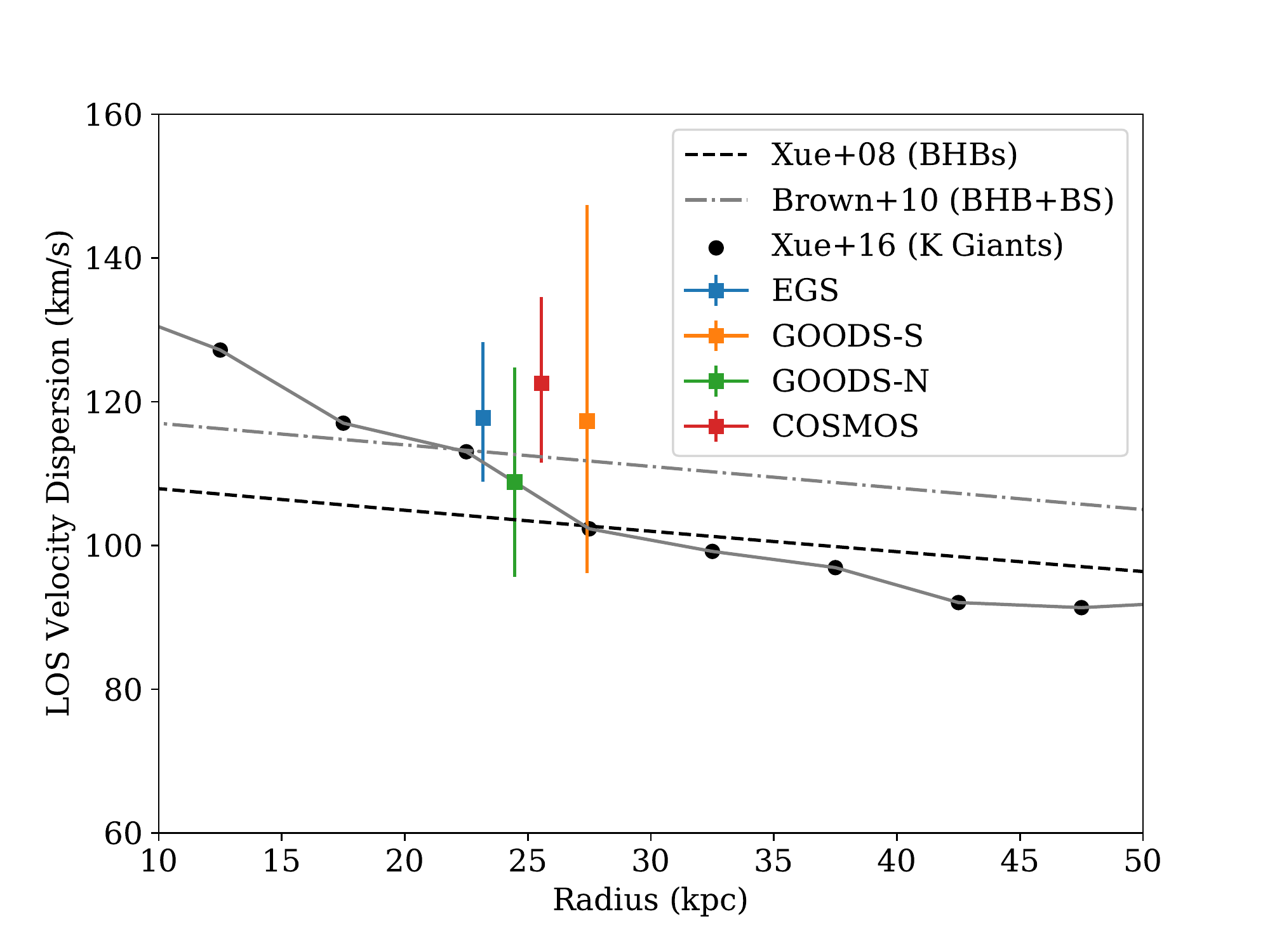}
	\caption{LOS velocity dispersions of the four HALO7D fields, plotted as a function of mean Galactocentric radius. Vertical errorbars show the 16-84\% quantiles of the marginalized posterior. We compare our LOS dispersions with results from other studies: the black dashed line indicates best-fit LOS dispersion profile from \cite{Xue2008}, measured from BHBs in SDSS. The grey dashed line indicates the best-fit dispersion profile from \cite{Brown2010} study, using BHB and BS stars as tracers. The black connected points show the resulting dispersion profile from the SEGUE K-giant survey \citealt{Xue2016}. The HALO7D dispersions are consistent with predictions from other tracers.}
	\label{fig:xue_h7d}
\end{figure}

	Figure \ref{fig:xue_h7d} shows the LOS dispersions of the four HALO7D fields plotted as a function of mean galactocentric radius (where we have converted heliocentric distance $\langle D \rangle$ to Galactocentric radius $\langle r \rangle$). Points indicate the median of the $\sigma_{LOS}$ posterior samples, and error bars indicate the 16 and 84 posterior percentiles. We compare our results to the velocity dispersion profiles measured in other studies. The figure also shows measured velocity dispersion profiles from SDSS blue horizontal branch (BHB) stars (\citealt{Xue2008}, black dashed line); BHB and blue straggler (BS) stars from the Hypervelocity Star Survey (\citealt{Brown2010}; grey dashed line); and the SEGUE K-giant survey (\citealt{Xue2016}; connected black dots). 
    The measured LOS velocity dispersions in the HALO7D fields using MSTO stars are consistent with other studies that have measured the LOS velocity dispersion profile over our distance range. 
	
	\section{Comparison with Simulations}
	\label{sec:bj}
	
	\begin{figure*}
		\includegraphics[width=\textwidth]{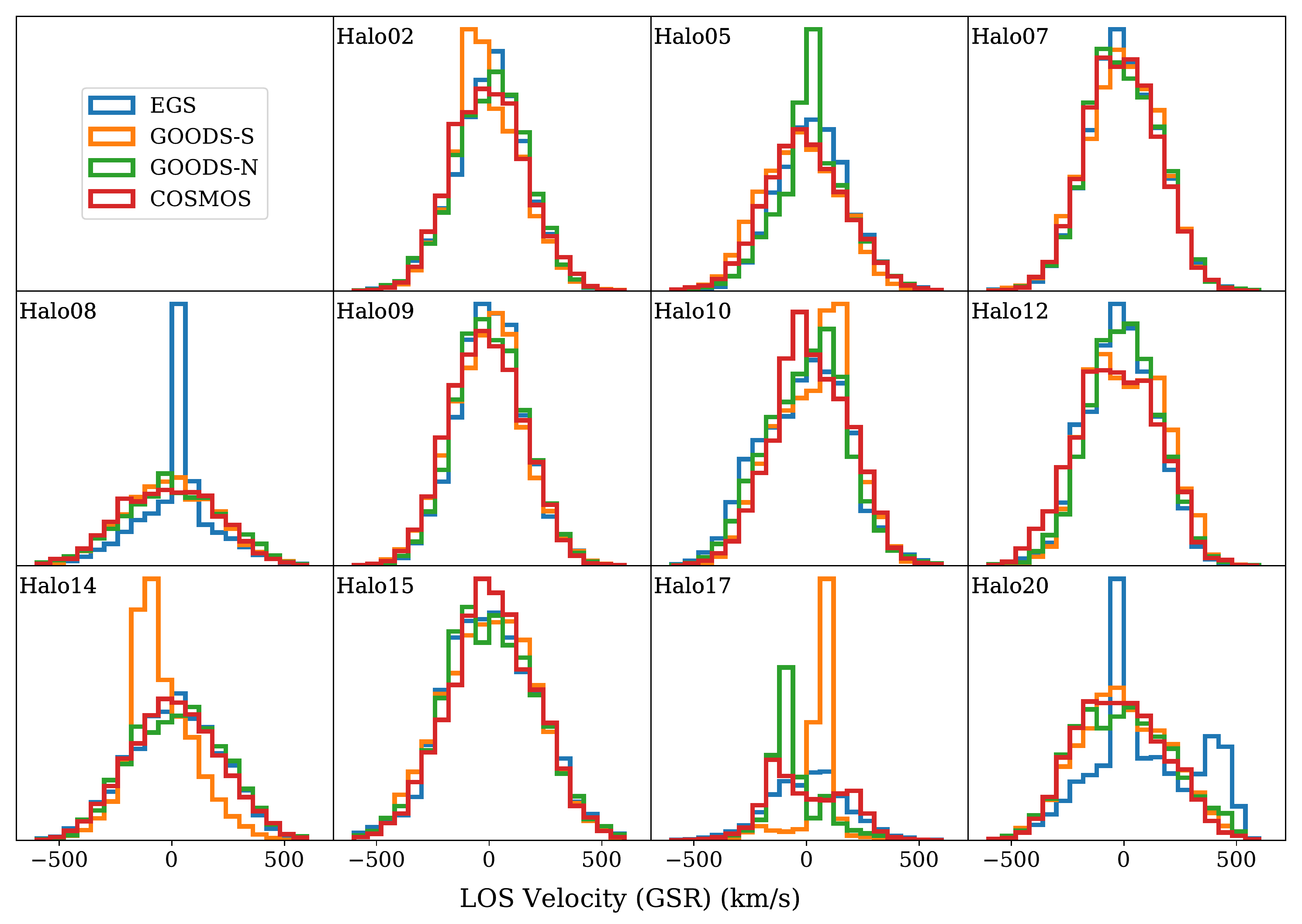}
		\caption{LOS velocity distributions for the mock HALO7D observations generated with Galaxia from the eleven BJ05 accreted stellar halos. Different colored histograms denote the observations in the different HALO7D fields. Seven out of the eleven BJ05 halos show consistent velocity distributions across the four fields. Three halos show consistency across three fields with one field dominated by substructure. Halo17 shows four distinct LOS distributions across the four fields.}
		\label{fig:bj_vlos}
	\end{figure*}

	\begin{figure*}
		\includegraphics[width=\textwidth]{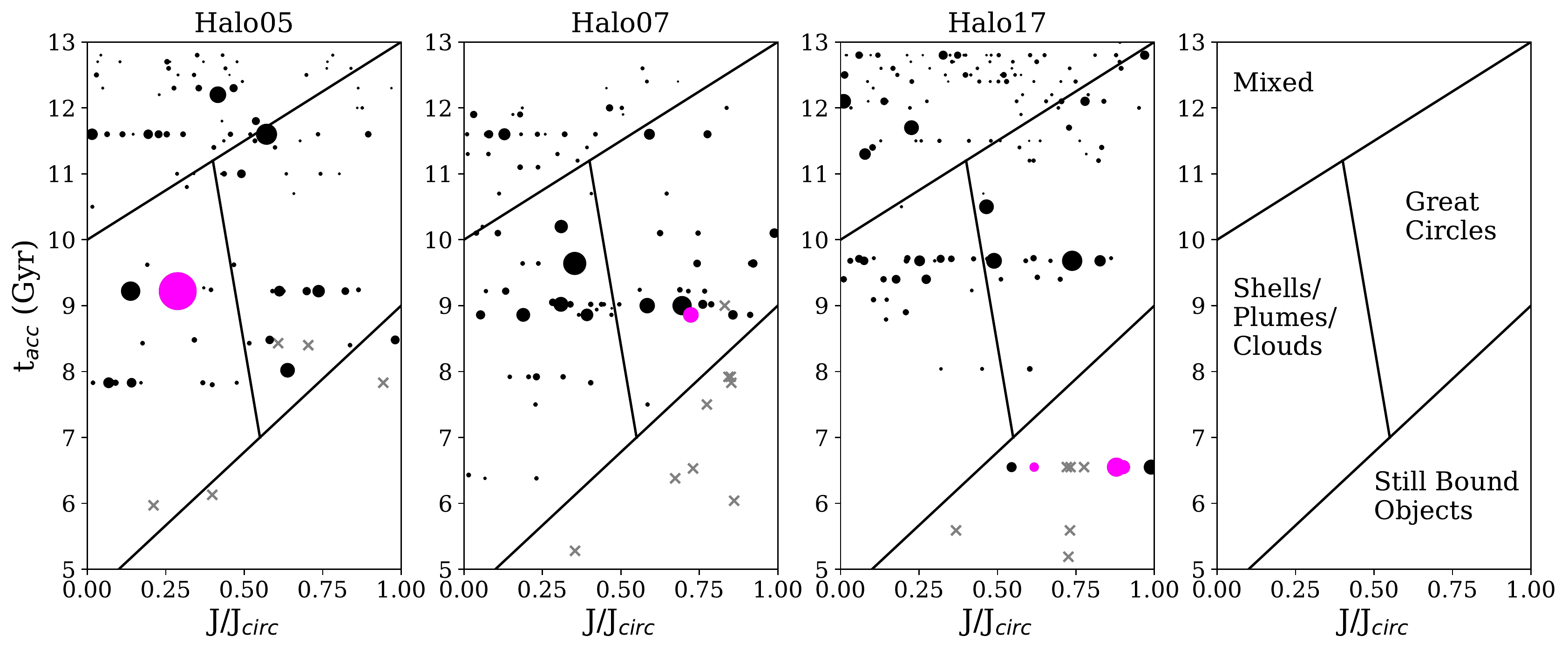}
		\caption{Age vs orbit circularity for the accretion events making up Halo05, Halo07 and Halo17. Points are scaled by accretion event mass. Grey crosses indicate still-bound satellites. Colored points indicate the ``dominant'' satellites in the mock HALO7D samples. For Halo05 and Halo07, the same satellite is dominant across all four fields; in Halo17, one satellite dominates two fields while the other two fields are dominated by distinct satellites. Halo17's epoch of recent accretion is responsible for the variable velocity distributions, while early, massive accretion results in consistent velocity distributions for Halo05 and Halo07. The far righthand panel is a reproduction of Figure 3 from \cite{Johnston2008}, showing the dominant morphological types that arise from accretion events in the different regions of this plane.}
		\label{fig:bj_tacc_jcirc}
	\end{figure*}
	
	\begin{figure}
		\includegraphics[width=0.5\textwidth]{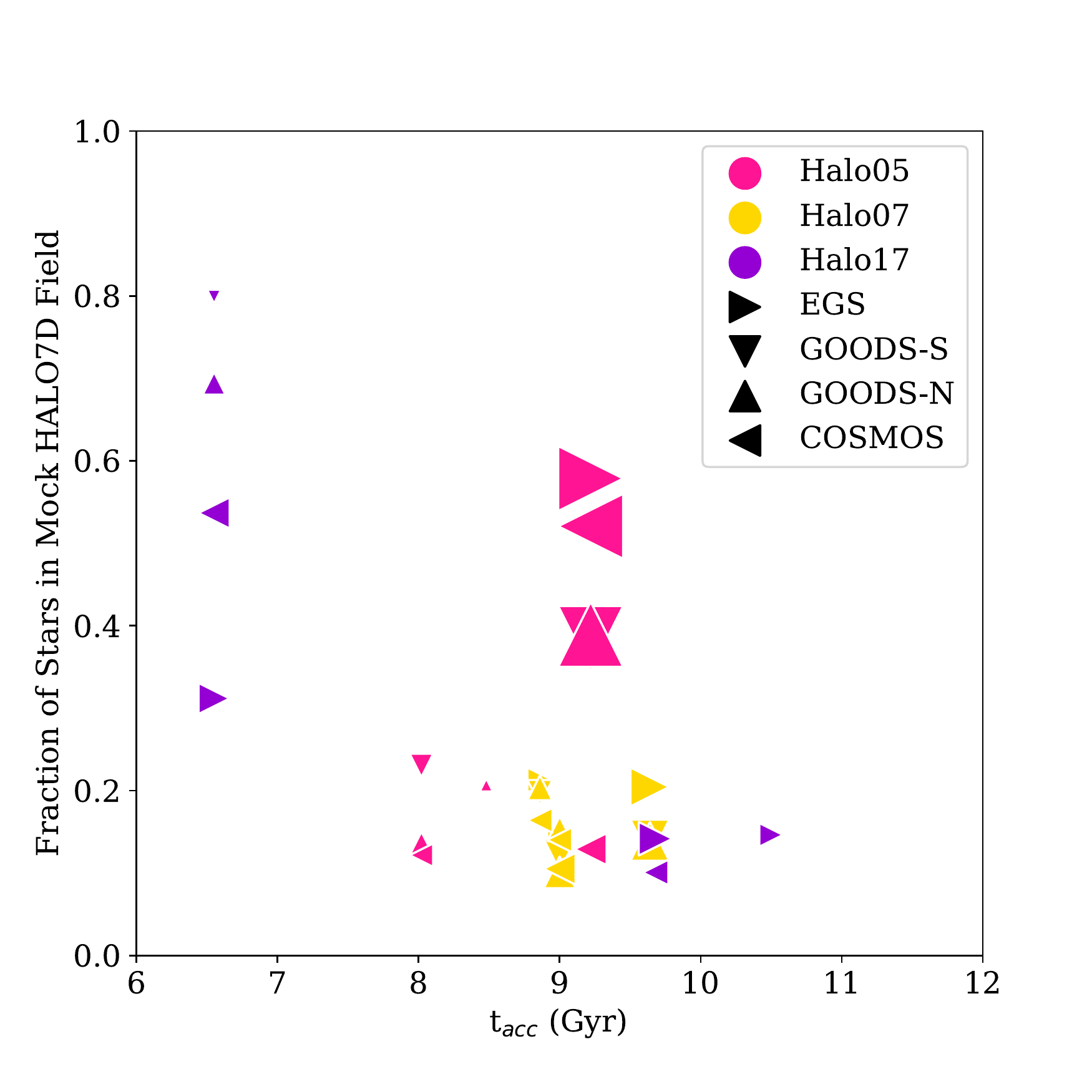}
		\caption{Fraction of stars contributed to a given line of sight as a function of accretion time, for Halo05 (magenta), Halo07 (gold) and Halo17 (purple). Point sizes are scaled by the mass of the accretion event. Only events that contribute $>10\%$ of the stars are shown. Different shape orientations denote different sightlines. Halo05's most massive satellite dominates the mock HALO7D sample along all sightlines, contributing 40-60\% of the stars. In Halo07, several relatively massive accretion events that were accreted around the same time all contribute between 10-20\% of the stars in the four sightlines. These two accretion histories give rise to consistent velocity distributions along the different sightlines. In contrast, Halo17 has experienced recent accretion of low mass satellites. These recent accretion events dominate the four HALO7D sightlines, and create cold peaks in the LOS velocity distributions.}
		\label{fig:bj_tacc_frac}
	\end{figure}
	
	In all four lines of sight, we see the ``hot halo'' population; none of our fields are dominated by substructure that is cold in LOS velocity. In addition, the measured LOS velocity dispersions across the four fields are all consistent with one another, and are consistent with measurements made using other tracers.
	
	In order to investigate whether or not this result is expected for a halo comprised primarily of accreted dwarfs, we performed mock HALO7D surveys on the eleven \cite{Bullock2005} halos (hereafter BJ05), using the synthetic survey software Galaxia (\citealt{Sharma2011}). The publicly available Bullock \& Johnston simulations are high resolution N-body simulations of accreted dwarf galaxies onto a Milky Way-like parent galaxy. The parent galaxy has a time-dependent analytical potential with halo, disk and bulge components. Because there is no stellar disk in these simulations, there is no ``in-situ'' stellar halo component in these galaxies. Galaxia can accept N-body simulations as input, and it generates synthetic catalogs with smooth, continuous distributions of stars over any given volume. 
    
    For our mock HALO7D surveys, we ``observed" one square degree areas centered on the coordinates of the four HALO7D fields in all eleven BJ05 halos. We choose to create catalogs of synthetic fields that are larger than our survey fields, because we are interested exploring the underlying LOS velocity distributions along these lines of sight; the larger area provides us with more samples from these distributions. We then selected stars that fell within the HALO7D CMD selection boxes.	

	Figure \ref{fig:bj_vlos} shows the LOS velocity distributions for the four HALO7D fields in each of the the eleven BJ05 halos. In seven out of the eleven halos, the four LOS velocity distributions are all ``hot", and consistent with one another. In three of the remaining halos, three of the fields have consistent LOS velocity distributions with one field having a strong cold peak (Halo08, Halo14, and Halo20). The only halo with four different velocity distributions across the four lines of sight is Halo17. 
	
	We now look at three halos more closely; Halo05, Halo07, and Halo17. The accretion histories of these halos are shown in Figure \ref{fig:bj_tacc_jcirc}, in the space of accretion time vs. $J_{sat}/J_{circ}$, with point sizes scaled by the mass of the accretion event. Lines indicate the regions of this plane dominated by the different morphological types discussed in \cite{Johnston2008}; this breakdown is shown in the far righthand panel of Figure \ref{fig:bj_tacc_jcirc} (see also their Figure 3). Figure \ref{fig:bj_tacc_frac} shows the accretion events that contribute $>10\%$ of the stars along a given sightline in the three halos; point sizes are again scaled by mass, with the different triangle orientations indicating the different sightlines. 
	
	Halo05 experienced a very massive ($4\times 10^{11} M_{\odot}$) accretion event around 9 Gyr ago; this event is highlighted by the colored point in the far lefthand panel of Figure \ref{fig:bj_tacc_jcirc}. This massive accretion event strongly dominates the mock HALO7D samples across all four sightlines, making up $40-60\%$ of the observed stars in this halo (pink points in Figure \ref{fig:bj_tacc_frac}). In four out of the eleven BJ05 halos, the most massive unbound satellite dominates the mock HALO7D samples in all four sightlines. 
	
	The HALO7D fields in Halo07 also have hot, consistent velocity distributions, though none of the sightlines are dominated by the debris from just one satellite.  Halo07 experienced several coincident accretion events of similar mass at $\sim 9$ Gyr. The three most massive of these events each contribute $10-20$ \% of the stars in each of the four fields (gold points in Figure \ref{fig:bj_tacc_frac}). These three, relatively massive, early, coincident events are responsible for the consistent, hot velocity distributions observed in Halo07.
	
	In contrast, Halo17's velocity distributions are not consistent across the four mock HALO7D fields. Figure \ref{fig:bj_tacc_jcirc} shows that Halo17's accretion history is characterized by several recent accretion events ($t_{acc}\sim 6.5$ Gyr) on fairly circular orbits; Figure \ref{fig:bj_tacc_frac} shows that these recent accretion events strongly dominate the HALO7D sightlines, with the most massive recent event dominating two sightlines ($f_{sample}\sim 30-50 \%$) and the other two sightlines are strongly dominated by two different less massive events ($f_{sample}=70-80\%$). The recent accretion experienced by this halo results in cold LOS velocity distributions along two of the four HALO7D sightlines.
	
	To summarize, early, massive accretion events (or several, early, synchronous accretion events) give rise to consistent, hot velocity distributions along different halo sightlines, whereas recent accretion can lead to sightlines dominated by kinematically cold substructure. While it is challenging to distinguish between the accretion histories of Halo05 and Halo07 with kinematics alone, we note that the chemical abundances will be different for these two scenarios. An accretion history like Halo05's should give rise to a higher average $[\mathrm{Fe/H}]$ than Halo07's accretion history, because of the mass-metallicity relation (e.g., \citealt{Kirby2013}).
	
	In their analysis of the BJ05 halo density profiles, \cite{Deason2013a} found that halos with early, massive accretion events had breaks in their density profiles (like the density profile of the MW; e.g., \citealt{Watkins2009}, \citealt{Deason2011b}, \citealt{Sesar2011}), whereas galaxies with prolonged accretion epochs had single power-law density profiles (like M31; e.g., \citealt{Gilbert2012}). Recent results from \textit{Gaia} have discovered the remnant of an early, massive accretion event, known as the ``Gaia-Sausage'' (\citealt{Belokurov2018}) or ``Gaia-Enceladus'' (\citealt{Helmi2018}), which is both relatively metal rich and strongly radially biased in its orbital distribution. \cite{Deason2018b} find that the apocenters of these ``Sausage'' stars are at $r\sim 20$ kpc, coincident with the MW's break radius; this is also the approximate mean distance to our sample. Studying the LOS velocity distributions of the simulated BJ05 halos, we find that the consistent LOS velocity distributions of the HALO7D fields provides yet another piece of evidence that the MW likely experienced a massive, early accretion event. Proper motion information and abundances will help us to determine if our sample is dominated by Gaia-Sausage stars. 
	
	\section{Conclusions}
	\label{sec:concl}
	
	In this paper, we presented the spectroscopic component of the HALO7D survey; a Keck II/DEIMOS spectroscopic survey of distant, MSTO MW halo stars in the CANDELS fields. We described the survey observing strategy, mask layouts, and target selection. We also presented a new method of measuring velocities from stellar spectra from multiple observations, utilizing Bayesian hierarchical modeling. We used the measured LOS velocities to estimate the parameters of the LOS velocity distributions in the four HALO7D fields.
	
	We summarize our conclusions as follows:
	\begin{enumerate}
    \item{When performing slit spectroscopy of point sources, it is essential to consider the apparent velocity shift due to slit miscentering when measuring velocities from individual spectra or when combining multiple spectroscopic observations. The hierarchical Bayesian approach presented in this work (implemented in \textsc{Velociraptor}) allows for the parameters of individual observations to be modeled simultaneously, leveraging the available signal while properly propagating uncertainties.} 
		\item{All four HALO7D fields are dominated by the ``hot" halo population, and have consistent LOS velocity distributions. The estimates of the velocity dispersions are consistent with estimates derived using other tracer populations.}
		\item{We performed mock HALO7D observations using the synthetic survey software Galaxia to observe the \cite{Bullock2005} halos. We found that an early, massive accretion event (or several early events) can cause consistent, hot velocity distributions along the different sightlines. This consistency in the velocity distributions arises because the same satellite (or the same few satellites) dominate the halo population along all sightlines. The consistent HALO7D LOS distributions therefore could indicate that the MW experienced an early, massive accretion event (or perhaps several events), whose stars are dominating the samples of all four fields.}
	\end{enumerate}
	
	This paper is the first in the HALO7D series; our spectroscopy and the multi-epoch \textit{HST} imaging will enable us to measure proper motions and abundances for these same stars. HALO7D is a deep complement to the \textit{Gaia} mission: these stars in this dataset will be the faintest stars with measured 3D kinematics until LSST. With upcoming proper motions and abundances, we can continue to use the HALO7D dataset to improve our understanding of the Galaxy's formation. 
	
\section*{Acknowledgments}
	Over the course of this work, E.C. was supported by a National Science Foundation Graduate Research Fellowship, as well as NSF grant AST-1616540.  Partial support for this work was provided
by NASA through grants for program AR-13272 from the
Space Telescope Science Institute (STScI), which is operated
by the Association of Universities for Research in Astronomy
(AURA), Inc., under NASA contract NAS5-26555. E.C. expresses her profound gratitude to Alexander Rudy, a conversation with whom sparked the radial velocity measurement technique outlined in this work. E.C. also thanks Tony Sohn for his help with Figure \ref{fig:footprints}. A.D. is supported by a Royal Society University Research Fellowship. 
A.D. also acknowledges support from the STFC grant ST/P000451/1. P.G. and E.T. acknowledge support from the NSF grants AST-1010039 and AST-1412504. We thank the outstanding team at Keck Observatory for assisting us in our observations. This research made use of Astropy, a community-developed core Python package for Astronomy (Astropy Collaboration, 2013). This work has made use of the Rainbow Cosmological Surveys Database, which is operated by the Universidad Complutense de Madrid (UCM), partnered with the University of California Observatories at Santa Cruz (UCO/Lick,UCSC). We recognize and acknowledge the significant cultural role and reverence that the summit of Mauna Kea has always had within the indigenous Hawaiian community. We are most fortunate to have the opportunity to conduct observations from this mountain.
	

\appendix
\section{Testing Velociraptor}
In this appendix, we discuss the testing of the Velociraptor technique described in Section \ref{sec:vel} on fake data. To create fake spectra for testing, we degraded our template spectrum to a variety of signal to noise ratios. In order to realistically reproduce the noise due to the sky background for one of the HALO7D exposures, we took the noise array for an extremely faint extragalactic target that had no visible continuum or emission lines for a 20 minute exposure. 

\begin{figure}
\centering
\includegraphics[width=0.5\textwidth]{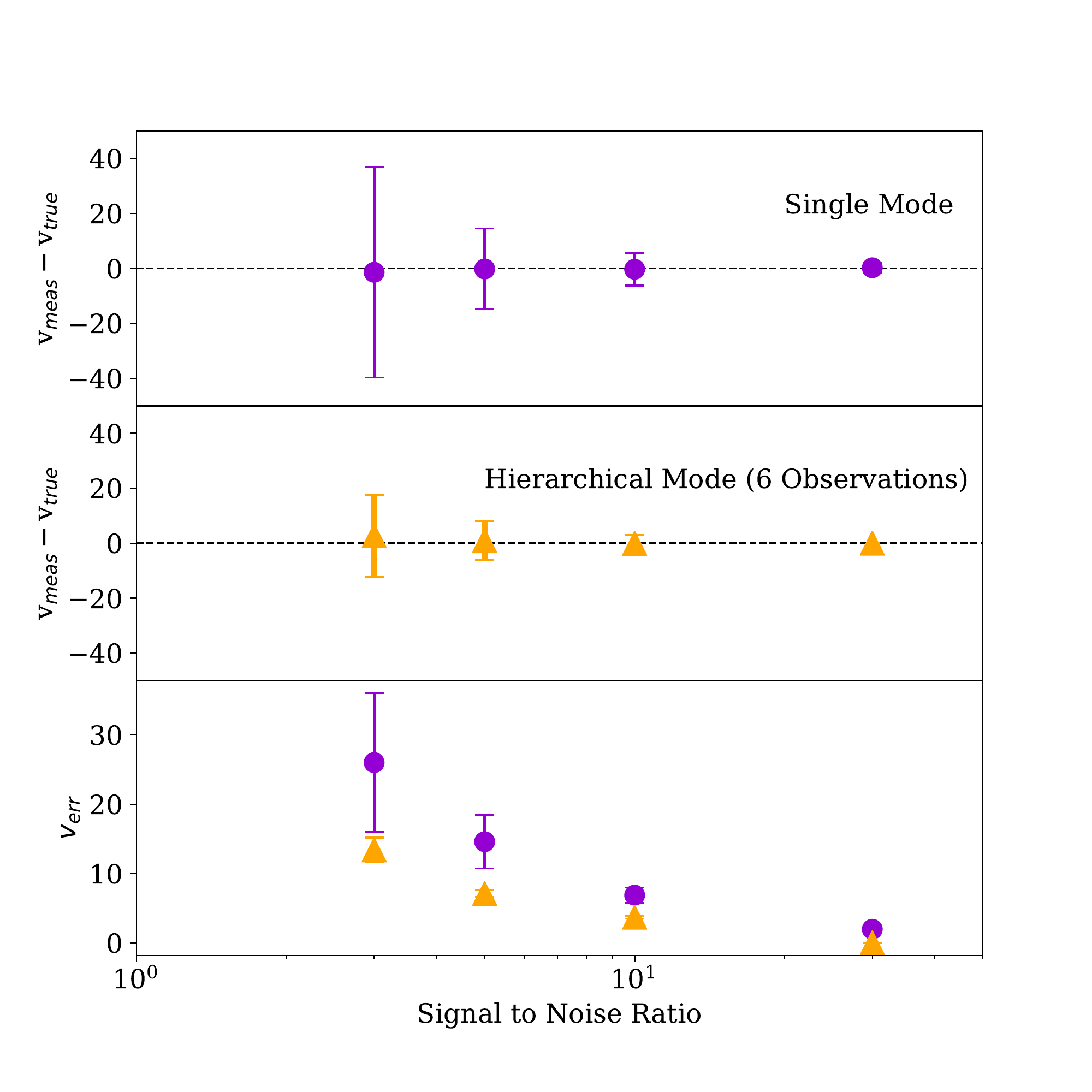}
\caption{Results from testing Velociraptor on fake data. Top panel: errorbars show the mean recovered velocity, and the standard deviation of the recovered velocities, for 90 runs of Velociraptor in single-mode, as a function of signal to noise. Middle panel: resulting distributions of recovered velocities when Velociraptor is run with six observations in hierarchical mode. Note that here the x-axis refers to the signal to noise of a single observation. Lower panel: velocity error (computed as half the difference between the 84th and 16th percentiles) in single-mode (purple) and hierarchical mode with six observations (orange). }
\label{fig:fake_sum}
\end{figure}

For signal to noise ratios of 3, 5, 10, and 30, we generated 180 individual spectra for each S/N. We tested Velociraptor in ``single mode" (i.e., working with individual observations only) for 90 of these spectra for each S/N. The mean recovered velocities, and the standard deviations of these velocities, are shown in the top panel of Figure \ref{fig:fake_sum}. For the spectra with S/N=3, approximately one third of the fake sample had failed velocity measurements (i.e., chains did not successfully converge); the errorbars in Figure \ref{fig:fake_sum} reflect the statistics for the successful measurements.

\begin{figure}
\includegraphics[width=\textwidth]{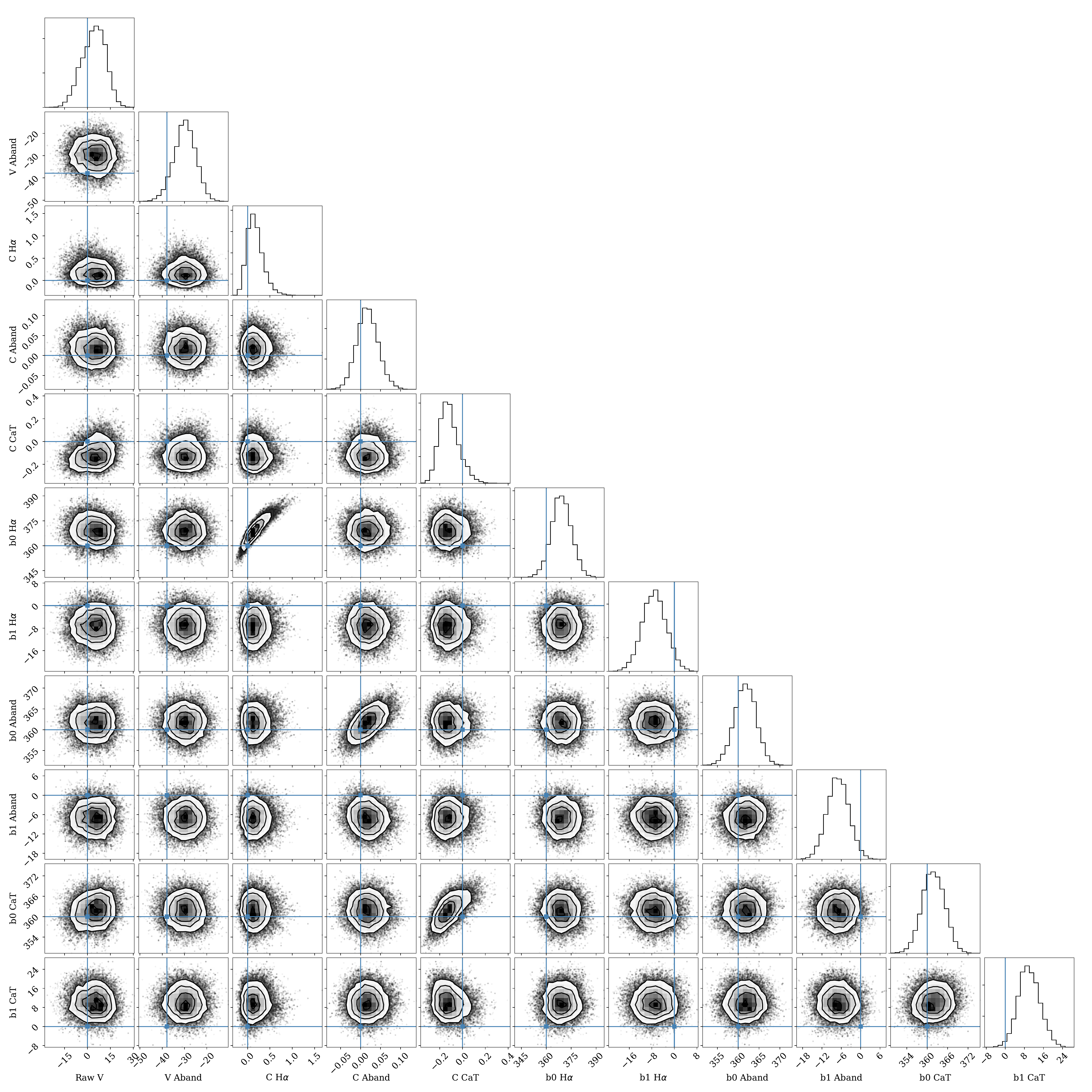}
\caption{Full corner plot for all 11 parameters for a fake spectrum with S/N=10. The true parameter values used to generate the fake spectrum are shown in blue. The absorption line strength parameters for a given spectral region are covariant with the continuum level; this is expected based on how we have parameterized the absorption lines.}
\label{fig:corner_fake}
\end{figure}

All projections of the posterior for one of our fake spectra can be seen in the corner plot in Figure \ref{fig:corner_fake}. The input values for the fake spectrum are shown as blue lines; the model successfully recovers the parameters of the fake data. The absorption line coefficients $C$ and the continuum levels for a given spectral region are covariant; this is expected because of the way in which we parametrized the absorption line strength (see Equation \ref{eqn:model_line}). 

\begin{figure}
\includegraphics[width=\textwidth]{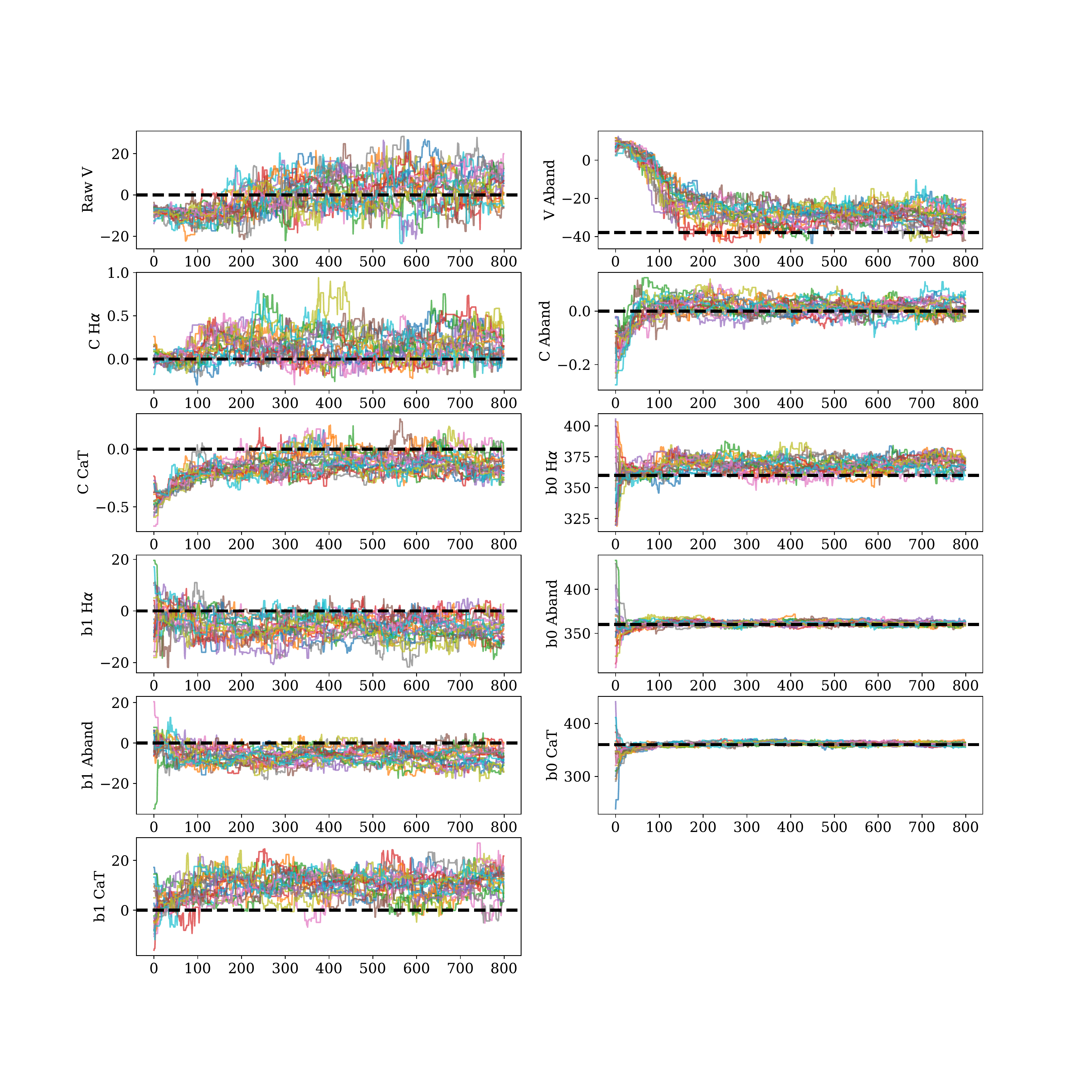}
\caption{Traces for all 11 single-mode parameters, for a fake spectrum generated to have S/N=10. For clarity, we show traces for only 20 randomly selected walkers. Black dashed lines indicate the true values of the model parameters used to generate this fake spectrum.}
\label{fig:trace_fake}
\end{figure}

Trace plots for 20 of the \verb+emcee+ walkers for the 11 parameters of single mode are shown in Figure \ref{fig:trace_fake}. The true parameter values that were used to generate the fake data are shown as thick black dashed lines. The traces are well mixed and converge successfully over the runtime of the sampler. 

To test Velociraptor in hierarchical mode, we combined six fake spectra at a given signal to noise ratio, and ran Velociraptor 30 times at each S/N. The resulting mean recovered velocities, and their standard deviations, are shown in the middle panel of Figure \ref{fig:fake_sum}, where here the signal to noise ratio plotted on the $x-$axis refers to the signal to noise of the individual observations. 

\begin{figure}
\includegraphics[width=\textwidth]{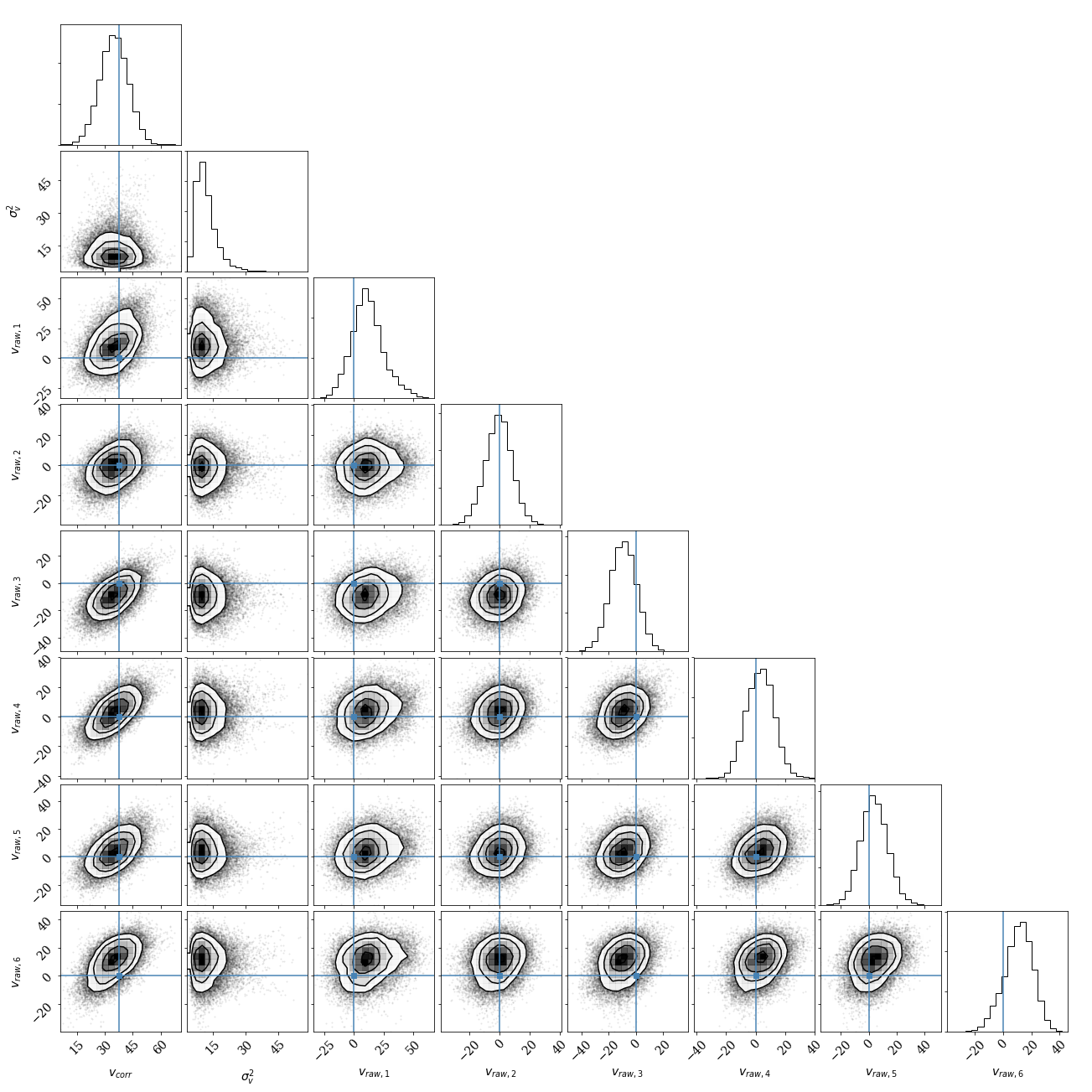}
\caption{Corner plot for the corrected velocity, the additional uncertainty $\sigma_v^2$, and the six raw velocities for six fake spectra that each have S/N=5. Note that we are only showing projections here for 8 out of the 68 parameters in this model. This particular run of Velociraptor ran for with 800 walkers for 500 steps. For this figure, we excluded the first 3000 steps as burn-in, and thinned the chain, showing every 50th sample.}
\label{fig:corner_hier}
\end{figure}

\begin{figure}
\includegraphics[width=\textwidth]{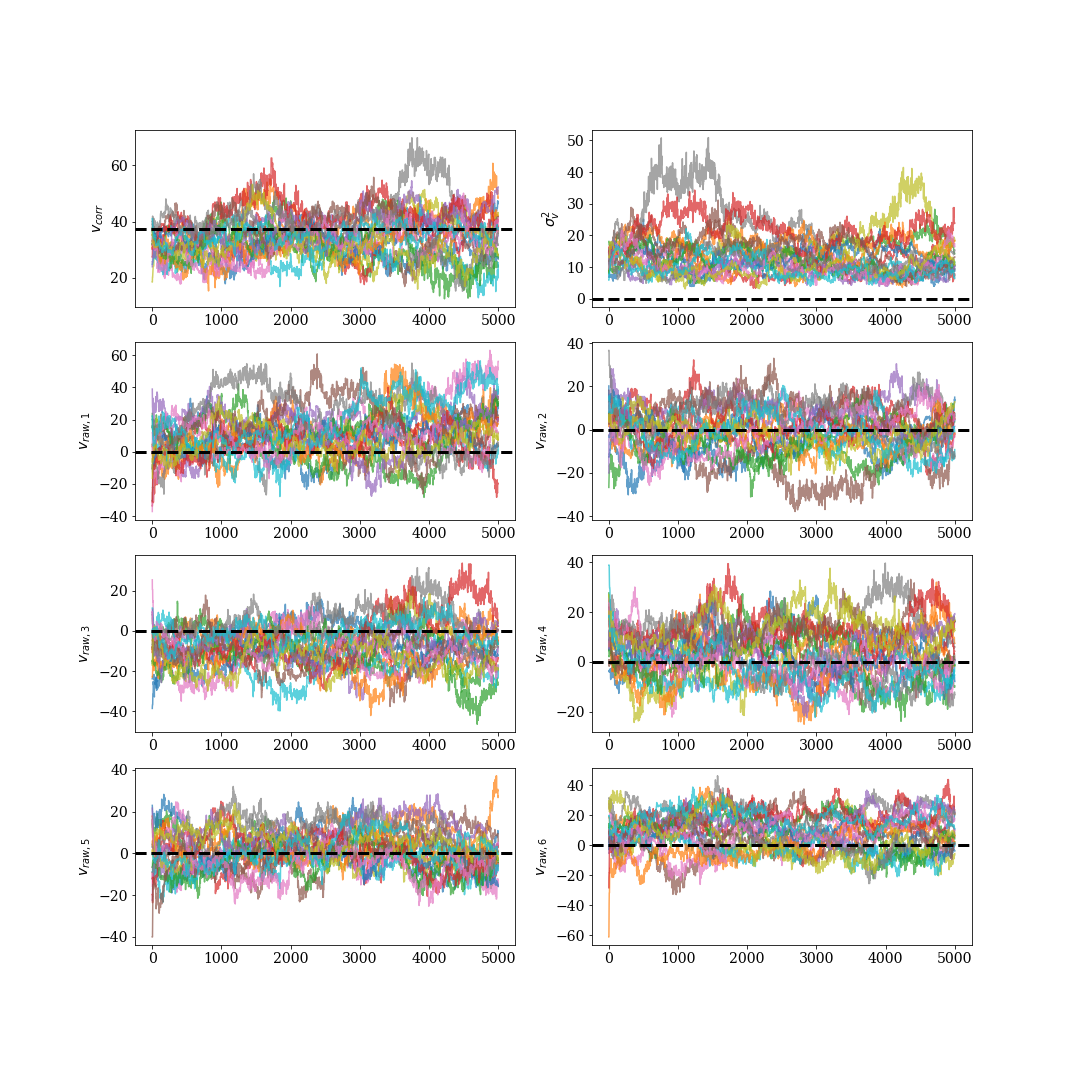}
\caption{Traces for the corrected velocity, the additional uncertainty $\sigma_v^2$, and the six raw velocities for six fake spectra that each have S/N=5. For clarity, we show traces for only 20 randomly selected walkers. Truths are shown as black dashed lines. Because of the complexity of the model and the large number of free parameters, the chains do not mix efficiently, and the sampler needs to be run for many iterations. Note that the true value for $\sigma_v^2$ is not recovered in this case, because all of our fake spectra were generated to have exactly the same velocity.}
\label{fig:trace_hier}
\end{figure}

Trace plots and a corner plot for the corrected velocity, the variance of velocities $\sigma_v^2$, and the raw velocity parameters for six observations, each with S/N=5, are shown in Figures \ref{fig:trace_hier} and \ref{fig:corner_hier}. We run \verb+emcee+ for 5,000 steps, and discard the first 3,000 as burn in. For the purposes of displaying the corner plot, we have ``thinned'' our chain, including every 50th sample for each walker. 

\end{document}